\documentclass[conference,compsoc]{IEEEtran}
\IEEEoverridecommandlockouts
\usepackage{amssymb}
\usepackage{graphicx}
\usepackage{epstopdf}
\usepackage{gensymb}
\usepackage{color}
\usepackage{amsmath}
\usepackage{bm}
\usepackage{etoolbox}

\usepackage{cite}
\usepackage{array}
\usepackage{booktabs}
\usepackage{multirow}
\usepackage{comment}
\usepackage{subcaption}
\usepackage{algorithm}
\usepackage{algorithmicx}
\usepackage{mathrsfs}
\usepackage{algpseudocode}

\usepackage{xcolor}
\usepackage[colorlinks,linkcolor={}]{hyperref}
\hypersetup{
  colorlinks,
  citecolor=black,
  linkcolor=black,
  urlcolor=urlblue}
\PassOptionsToPackage{hyphens}{url}\usepackage{hyperref}
\usepackage{enumitem}
\usepackage{wasysym}

\usepackage{balance}
\usepackage{soul}
\usepackage{xcolor}

\usepackage[framemethod=TikZ]{mdframed}
 
\alglanguage{pseudocode}

\usepackage{threeparttable}
\usepackage{mathtools}
\usepackage{hyperref}
\usepackage{pifont}
\usepackage{listings}

\usepackage{url}
\urldef{\mailsa}\path|{alfred.hofmann, ursula.barth, ingrid.haas, frank.holzwarth,|
	\urldef{\mailsb}\path|anna.kramer, leonie.kunz, christine.reiss, nicole.sator,|
	\urldef{\mailsc}\path|erika.siebert-cole, peter.strasser, lncs}@springer.com|    
\newcommand{\keywords}[1]{\par\addvspace\baselineskip
	\noindent\keywordname\enspace\ignorespaces#1}
\usepackage[utf8]{inputenc}
\usepackage[english]{babel}

\usepackage{amsthm}
\usepackage{balance}
\usepackage{xspace}

\usepackage{lipsum}
\usepackage{multicol}

\usepackage{bbding}

\usepackage{enumitem}

\theoremstyle{definition}
\newtheorem{definition}{Definition}
\definecolor{R}{RGB}{0,0,150}

\theoremstyle{remark}

\definecolor{myblue}{rgb}{0,0,0.9}

\definecolor{urlblue}{RGB}{60,132,196}
\definecolor{red}{RGB}{207,78,56}
\definecolor{gray}{RGB}{146,146,161}

\newcommand{\metricname}{exclusivity\xspace}
\newcommand{\backdoorname}{\textsc{BELT}\xspace}
\newcommand{\backdoorallname}{Backdoor Exclusivity LifTing\xspace}

\newcommand{\eat}[1]{}

\usepackage{fancyhdr}
\fancypagestyle{firstpage}{
  \fancyhf{}
  
  \fancyhead[C]{\vspace{10pt}\normalsize{To Appear in the 45th IEEE Symposium on Security and Privacy, May 20-23, 2024.}\\\vspace{-25pt}} 
  \fancyfoot[C]{\thepage}
}

\begin{document}

\title{BELT: Old-School Backdoor Attacks can Evade the State-of-the-Art Defense \\ with Backdoor Exclusivity Lifting}

\author{
   \IEEEauthorblockN{
       Huming Qiu,
       Junjie Sun,
       Mi Zhang\IEEEauthorrefmark{1},
       Xudong Pan,
       Min Yang\IEEEauthorrefmark{1}
   }
   \thanks{\IEEEauthorrefmark{1} Corresponding authors: Mi Zhang and Min Yang}

   \IEEEauthorblockA{
       Fudan University, China
   } 

   \IEEEauthorblockA{
       \{hmqiu23@m.,
       jjsun22@m.,
       mi\_zhang@,
       xdpan@,
       m\_yang@\}fudan.edu.cn
   } 
   
}

\maketitle

\thispagestyle{firstpage}

\begin{abstract}
Deep neural networks (DNNs) are susceptible to backdoor attacks, where malicious functionality is embedded to allow attackers to trigger incorrect classifications. Old-school backdoor attacks use strong trigger features that can easily be learned by victim models. Despite robustness against input variation, the robustness however increases the likelihood of unintentional trigger activations. This leaves traces to existing defenses, which find approximate replacements for the original triggers that can activate the backdoor without being identical to the original trigger via, e.g., reverse engineering and sample overlay.

In this paper, we propose and investigate a new characteristic of backdoor attacks, namely, \textit{backdoor \metricname}, which measures the ability of backdoor triggers to remain effective in the presence of input variation. 
Building upon the concept of backdoor \metricname, we propose \backdoorallname (\backdoorname), a novel technique which suppresses the association between the backdoor and fuzzy triggers to enhance backdoor \metricname for defense evasion.
Extensive evaluation on three popular backdoor benchmarks validate, our approach substantially enhances the stealthiness of four old-school backdoor attacks, which, after backdoor exclusivity lifting, is able to evade \textcolor{black}{seven} state-of-the-art backdoor countermeasures, at almost no cost of the attack success rate and normal utility. For example, one of the earliest backdoor attacks BadNet, enhanced by \backdoorname, evades most of the state-of-the-art defenses including ABS and MOTH which would otherwise recognize the backdoored model.
\end{abstract}

\section{Introduction}\label{sec:Intro}
Deep neural networks (DNN) are widely applied in mission-critical scenarios including face recognition \cite{sarkar2020facehack, wenger2021backdoor, parkhi2015deep},  autonomous
driving vehicles\cite{cao2021invisible, peng2020first}, and intelligent healthcare \cite{feng2022fiba}. 
Instead of training one's own DNN, increasingly more users are now integrating third-party models from popular platforms like HuggingFace in their applications (e.g., the daily downloads of Huggingface even surpasses those of traditional software supply chains including NPM and Pypi \cite{jiang2022empirical}). However, DNNs can be infected by \textit{backdoor attacks}.
The objective of backdoor attacks is to embed malicious functionality into a DNN model in a stealthy way, which would then incorrectly classify samples carrying \textit{triggers} (i.e., from pixel patterns to physical objects) as target labels. Meanwhile, the prediction remains intact on normal data \cite{li2022backdoor}. In the example of face recognition models, a backdoored model would allow unauthorized access of a person with a special pair of eyeglasses \cite{chen2017targeted}. Consequently, backdoor attacks are considered one of the severest threats against deep learning systems \cite{gao2020backdoor, baracaldo2022machine}.

\begin{figure}
    \centering
    \includegraphics[trim=0 0 0 0,clip,width=0.35\textwidth]{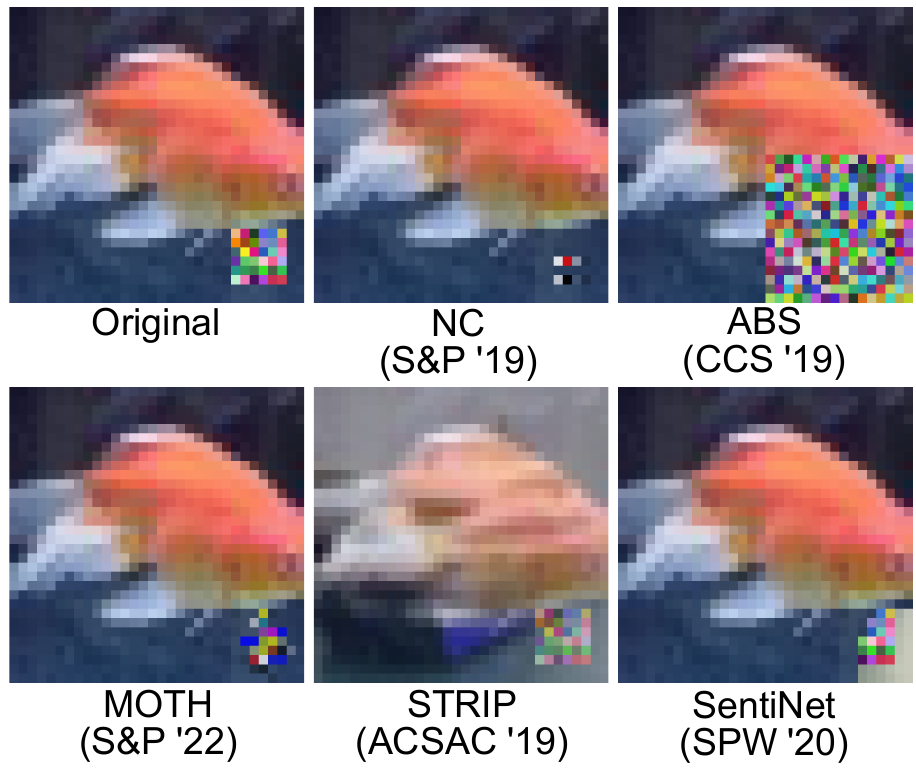}
        \caption{Examples of fuzzy triggers employed by backdoor countermeasures on classical backdoor attacks.}
    \label{fig:defense_trigger}
\end{figure}

In existing literature \cite{xue2021robust, zhang2022poison, wu2022just, xue2023compression, weber2023rab}, \textit{backdoor robustness} (i.e., the ability of backdoor triggers to remain effective in the presence of variations or noise) arises as an important property for most backdoor attacks. In fact, existing attacks require the backdoor can be activated by a trigger under a variety of input conditions. Therefore, attackers tend to design triggers of \textit{strong features}, such as random pixel patches \cite{gu2017badnets} or fixed patterns \cite{chen2017targeted}, which can be easily learned by the victim model. In contrast to complex semantic features, trigger features are easier for the model to learn \cite{li2021anti, yang2022not}. This provides backdoor attacks with two significant advantages.
On the one hand, it propels the model to exhibit stronger generalization capabilities for trigger features, enabling backdoor attacks to achieve nearly $100\%$ attack success rates (ASR). On the other hand, strong features remain recognizable under input noise or interference, which enhances the backdoor robustness. 
Therefore, backdoor robustness is a natural advantage, increasing the success rate and stability of backdoor attacks, allowing triggers to remain effective in complex environments. 

However, backdoor robustness also results in a numerous number of  approximate triggers that can unintentionally activate the backdoor. This increases the possibility of the backdoor to be detected by a model inspector. Most of the state-of-the-art  backdoor defenses \cite{gao2019strip,liu2019abs,wang2019neural,chou2020sentinet,xu2021detecting,tao2022model} do transform the robustness advantage of the backdoor into a weakness for trigger proximity. 
A prerequisite for the successful implementation of these backdoor defenses is the ability to effectively find an approximate replacement for the original trigger, leveraging methods such as reverse engineering \cite{wang2019neural}, sample overlay \cite{gao2019strip}, or attribution techniques \cite{chou2020sentinet}.
For instance, MOTH published at S\&P 2022 \cite{tao2022model} achieves model orthogonalization by relearning reverse triggers and natural triggers, which \textcolor{black}{results in} effective backdoor elimination. Section \ref{methodology_description} provides more detailed analysis. As shown in~\autoref{fig:defense_trigger}, despite the triggers identified by these defense strategies not being identical to the original trigger, they still possess the ability to activate the backdoor and can thus be classified as fuzzy triggers. 

\noindent{\bf Our Work.}  From our perspective, backdoor robustness is akin to the head of a coin, describing the tolerance of the backdoor to trigger variations, while the tail of the coin represents non-robustness, reflecting the precision of the backdoor's response to triggers.
In the following sections, we refer to non-robustness as \textit{backdoor \metricname} and term triggers that differ from the original triggers but can still activate the backdoor as \textit{fuzzy triggers}, with a formal definition provided in~\autoref{sec:prodef}. Given the prevalence of fuzzy triggers and their significant success in backdoor defense systems, a natural question arises: a) \textit{Can we measure the backdoor \metricname} of existing attacks? b) \textit{If so, can we devise approaches to suppress the existence of fuzzy triggers to evade the state-of-the-art defenses?} 

In this paper, we provide an affirmative answer by unveiling a novel \backdoorallname (\backdoorname) attack. \backdoorname is effective, easy to implement, and can synergize with existing state-of-the-art (SOTA) backdoor attacks to further enhance the escapability of the backdoor. We delve into exploring the close interplay between fuzzy triggers and backdoor \metricname, designing a universal and efficient measure of backdoor \metricname. Specifically, we conduct a comprehensive perturbation analysis of the original triggers, estimating the perturbation boundary that renders the triggers ineffective. This boundary indicates the potential range of existence for fuzzy triggers, demonstrating the maximum distortion that triggers can tolerate while still activating the backdoor. Leveraging this boundary, we introduce, for the first time, a metric for measuring backdoor \metricname. Additionally, we measure the \metricname of existing SOTA backdoors based on this metric, revealing that conventional backdoor attacks perform poorly in terms of \metricname, subsequently affecting their escape capabilities against countermeasures. This provides powerful insights for further research and improvement of backdoor attacks.

To this end, guided by backdoor \metricname, we propose a new backdoor attack technique \backdoorname. \backdoorname aims to precisely define the activation conditions of the backdoor, rendering fuzzy triggers incapable of activating the backdoor, thereby enhancing backdoor \metricname. Specifically, during the data poisoning phase, we create two batches of poisoned samples. One batch comprises dirty samples carrying the original triggers, with labels modified to the target label, aiming to inject the backdoor into the model. The other batch consists of covering samples carrying fuzzy triggers while still maintaining the ground truth labels, intending to suppress the association between the backdoor and fuzzy triggers, thereby strengthening backdoor \metricname. Finally, when the model is trained on this dataset containing these poisoned samples, a highly exclusive backdoor is implanted.

\vspace{1pt} \noindent{\bf Our Contributions.} In summary, we mainly make the following key contributions:

\noindent$\bullet$ We propose and investigate a new characteristic of backdoor attacks, i.e., \textit{backdoor \metricname}, which measures the ability of backdoor triggers to remain effective in the presence of input variation. By devising an algorithm for quantifying backdoor \metricname, it provides a new and practical lens to the stealthiness of existing backdoor attacks.

\noindent$\bullet$ On the basis of backdoor \metricname, we propose a new backdoor attack technique called \backdoorname, which suppresses the association between the backdoor and fuzzy triggers to enhance backdoor \metricname for defense evasion. Our technique can be \textcolor{black}{seamlessly} combined with existing backdoor attacks and enhance the classical attacks to evade the SOTA backdoor defenses which would otherwise detect and eliminate them.

\noindent$\bullet$ We comprehensively evaluate the performance of \backdoorname on \textcolor{black}{seven} SOTA backdoor defenses, and the results validate that \backdoorname helps four classical backdoor attacks to evade the SOTA backdoor defenses, while preserving the attack success rate and the clean accuracy. This confirms that highly exclusive backdoors are more difficult to be detected.
To facilitate future studies, we open-source our code in the following repository: \url{https://github.com/JSun20220909/BELT}.

\section{Background}\label{sec:bkgd}

\subsection{Backdoor Attacks}
Backdoor attacks aim to implant a backdoor into DNN (Deep Neural Network) models, establishing a secret mapping relationship between triggers and target labels. This allows the model to behave normally on benign samples but exhibit anomalous behavior on samples containing triggers. In simple terms, triggers carried by the inputs and the backdoor embedded in the model are the two crucial elements necessary for the successful execution of a backdoor attack \cite{gao2020backdoor}. The concealed backdoor in the model responds to triggers applied to the input, activating the backdoor effect, leading the model to execute specific malicious actions, such as misclassifying samples into designated categories. 

The development of backdoor attacks is primarily focused on enhancing the stealthiness or evasiveness of the backdoor \cite{li2020backdoor}. Two main development paths revolve around designing 1) special trigger types and 2) advanced backdoor types. For the former, the most traditional triggers include patches with fixed positions and patterns \cite{gu2017badnets}, as well as semi-transparent patterns blended with images \cite{chen2017targeted}. Later on, natural features from datasets themselves or external features have also been used as triggers \cite{liu2020reflection, ma2023horizontal}. Additionally, imperceptible triggers, such as subtle noise \cite{li2021invisible} and style transfer \cite{cheng2021deep} have been designed. Recently, research has explored composite triggers \cite{lin2020composite} and sample-specific triggers \cite{nguyen2020input}.

Concerning the latter, the most common backdoor type is the source-class-agnostic backdoor \cite{gao2020backdoor, li2023ntd}, where the backdoored model consistently misclassifies trigger samples into the specified labels. Another common type is the source-class-specific backdoor \cite{tang2021demon, wang2023cassock}, where the backdoored model only misclassifies trigger samples belonging to a specific source class, and the all-to-all attack falls into this category \cite{doan2021lira}. Furthermore, more advanced backdoor types have been proposed, such as the abuse of commercial quantization toolkits for quantization backdoors (specifically TensorFlow-Lite and PyTorch Mobile) \cite{ma2023quantization, pan2021understanding}.

\subsection{Robustness of DNN and Backdoor}
The robustness of a neural network refers to its stable predictive capability in the presence of noise, perturbations, or unknown variations in the input data \cite{gawlikowski2023survey}. It is commonly regarded as the model's tolerance to changes in the data. Similarly, the robustness of a backdoor pertains to the model's tolerance to variations in the trigger of the backdoor. In other words, it reflects the capability of the backdoor to maintain a stable response even when exposed to a slight change in the trigger. The robustness boundary of a model is specific to a particular sample and ensures the maximum perturbation range for which the model's predictions remain correct. Under the assumption of $L_p$ space, the model's $\epsilon$-robustness guarantee for $x$ indicates that the model's classification decision for $x$ will not change within an $\epsilon$ radius around this sample in the $L_p$ space.

Due to the presence of non-linear activation functions and the complex structure of deep neural networks, they exhibit non-linearity and non-convexity, making it challenging to estimate their output ranges. Computing precise robustness boundaries is an NP-complete problem \cite{katz2017reluplex}. As a result, many research focuses on employing approximation methods to compute model robustness boundaries \cite{weng2018evaluating, cohen2019certified, li2023sok}. These methods typically rely on robust optimization principles and estimate the model's robustness boundary by solving the inner maximization problem in ~\autoref{eq:robust_boundaries}:

\begin{equation}\label{eq:robust_boundaries}
\min _\theta \frac{1}{n} \sum_{i=1}^n \max _{\left\|x_i^{\prime}-x_i\right\|_p \leq \epsilon} \mathcal{L}\left(f_\theta\left(x_i^{\prime}\right), y_i\right).
\end{equation}

Here, $\mathcal{L}$ represents the loss function, $f_\theta$ represents the model, $x_i$ represents the original data sample, $y_i$ represents the label, $x_i'$ represents the adversarial sample, and $\epsilon$ represents the constraint on the perturbation magnitude.

\section{Problem Definition}
\label{sec:prodef}

Formally, a neural network $F$ consists of multiple layers of interconnected artificial neurons, with a parameter set denoted as $\theta$. It can be modeled as a non-linear mapping function from input data $X$ to output data $Y$, i.e., $F_\theta : X \rightarrow Y$.
For a classification task with $K$ categories, the mapping function is defined as $F_\theta : X \subset \mathbb{R}^d \rightarrow Y \subset \mathbb{R}^K$. Given an input sample $x \in X$, its prediction $c(F_\theta (x)) = \text{argmax}_{1 \leq i \leq K} F_\theta(x)_i$, where $F_\theta(x)_i$ is the $i$-th element of $F_\theta(x)$.

A backdoor attack entails the implantation of a backdoor into the neural network $F$, resulting in a modified mapping function denoted as $F_{\theta'}$, where $\theta'$ represents the compromised parameter set. For a clean input $x \in X$, the predictions of the backdoored model $F_{\theta'}$ always align with those of the clean model $F_\theta$, i.e., $c(F_{\theta'}(x)) = c(F_\theta(x)) = y$. For a poisoned input $x_t \in X$ carrying a trigger, the backdoored model $F_{\theta'}$ activates the backdoor effect, mapping it to the target class $y_t$ specified by the attacker, i.e., $c(F_{\theta'}(x_t)) = y_t$. We use $\mathbf{T}(\cdot)$ to denote the trigger transformation function, used to convert clean input into poisoned input. The trigger injection function has various forms~\cite{wang2019neural,li2020backdoor}, and it can be unified as:
\begin{equation}\label{eq:trigger}
\mathbf{T}(x, m_x, p, m_p) = x \cdot m_x + p \cdot m_p,
\end{equation}
where $m_x$ and $m_p \in [0, 1]^d$ are the input mask and trigger mask, respectively, and $p \in \mathbb{R}^d$ is the trigger pattern.
The trigger $t$ is jointly determined by the pattern and mask, and depending on the design of each attack, it can be represented as $t = p \cdot m_p$. For instance, in the case of BadNet attack \cite{gu2017badnets}, the trigger is a patch block with $m \in \{0, 1\}^d$ and $m_x = 1 - m_p$. Similarly, for the Blended attack \cite{chen2017targeted}, the trigger is a semi-transparent pattern background with $m \in [0, 1]^d$ determining the pattern's transparency, and $m_x = 1 - m_p$.

Intuitively, a backdoor is a hidden mapping function within deep learning models. Its purpose extends beyond inducing malicious behavior leading to misclassifications; it also aims to achieve concealment in backdoor detection and robustness in backdoor mitigation.
To achieve these goals, some attack methods focus on minimizing the $l_p$-norm of triggers \cite{li2021invisible}, i.e., $\text{argmin} \|t\|_p$, while others consider associating triggers with normal features \cite{lin2020composite}. All these trigger techniques are designed to evade countermeasures against backdoor defenses. However, these sophisticated techniques often overlook the \metricname of the backdoor in their design. In other words, the activation conditions of even well-designed triggers are not precise enough, thus making the backdoor susceptible to activation by triggers outside the intended design.

Next, we formally define these concepts. Consider a typical image classification problem where a sample $x \in [0, 1]^d$ and the corresponding label $y \in \{0, 1, \ldots, k\}$ follow the joint distribution $D(x, y)$. A neural network $F_\theta : [0, 1]^d \rightarrow \{0, 1, \ldots, n\}$ with parameters $\theta$ should satisfy the property $\arg\max_\theta \mathbb{P}(x,y) \sim D[F_\theta(x) = y]$.

\begin{definition}\label{def:one}
(Perturbed Trigger and Perturbation Boundary).
Given an original trigger $t \in \mathbb{R}^d$, if $t' = t + \delta$ and $\|\delta\|_p \neq 0$, we say $t'$ is a perturbed trigger of $t$ with perturbation $\delta \in \mathbb{R}^d$ and $l_p$-norm $\|\delta\|_p$. The perturbation boundary $\delta_b$ for $t$ within the numerical allowance is expressed as  $\delta_b = t_{max} - \text{round}\left(\frac{{t - t_{min}}}{{t_{max} - t_{min}}}\right) \cdot (t_{max} - t_{min}) - t$.
\end{definition}

\begin{definition}\label{def:two}
(Fuzzy Trigger and Ineffective Trigger).
Given a backdoored model $F_{\theta'}$, an original trigger $t$, a perturbed trigger $t'$, and a target label $y_t$. If $c(F_{\theta'}(x_{t'})) = c(F_{\theta'}(x_t))$, we refer to $t'$ as an fuzzy trigger of $t$, i.e., a perturbed trigger capable of activating the backdoor. Conversely, if $c(F_{\theta'}(x)) \neq y_t \land c(F_{\theta'}(x_{t'})) \neq c(F_{\theta'}(x_t))$, we refer to $t'$ as an ineffective trigger, i.e., a perturbed trigger incapable of activating the backdoor.
\end{definition}

\begin{definition}\label{def:three}
(Trigger Upper Bound).
Suppose the maximum perturbation that a trigger can endure while retaining its ability to activate the backdoor is denoted as $\delta_{\text{max}}$, the upper bound for triggers, $\beta_U$, is defined as the norm of the maximum perturbation among all fuzzy triggers, expressed as $\beta_U = \|\delta_{\text{max}}\|_p$. For any perturbed trigger with $\|\delta\|_p > \beta_U$, it loses the capability to activate the backdoor.
\end{definition}

Specifically, for any perturbation $\delta \in \mathbb{R}^d$ with $\|\delta\|_p > \|\delta_{max}\|_p$, it holds that if $c(F_{\theta'}(x)) \neq y_t$ then $c(F_{\theta'}(x_{t'})) \neq c(F_{\theta'}(x_t))$. With a trigger upper bound $\|\delta_{max}\|_p$ for a poisoned input $x_t$, we can devise a metric $\text{Excl}$ to measure its \metricname. \textcolor{black}{Considering that different triggers have different number of pixels $N$ and range of values [$t_{min}$, $t_{max}$], \metricname can be expressed as:}
\begin{equation}\label{eq:sample_spe}
\textcolor{black}{\text{Excl} = 1 -  (\frac{\|\delta_{max}\|_p}{\|\delta_{b}\|_p})^{(t_{max}-t_{min})\sqrt{N}}.}
\end{equation}

\begin{figure}
    \centering
    \includegraphics[trim=0 0 0 0,clip,width=0.45\textwidth]{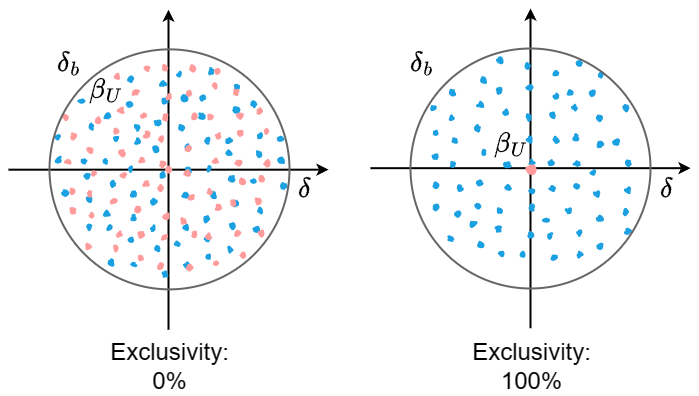}
        \caption{Schematic diagram of backdoor \metricname.}
    \label{fig:spec}
\end{figure}

In practice, we default to using the Euclidean norm ($l_2$-norm) for exclusivity. The \metricname metric $\text{Excl} \in [0, 1]$ is used to quantify the uniqueness of triggers. As shown in ~\autoref{fig:spec}, the origin represents the original trigger, the coordinate axis represents the size of the disturbance suffered by the original trigger, the black circle is the disturbance boundary $\delta_b$, \textcolor{black}{and the red and blue points respectively represent the fuzzy trigger and ineffective trigger.} For a backdoor with \metricname approaching 0\%, $\beta_U$ approaches the perturbation boundary $\delta_b$. In this case, there is no clear trigger upper bound $\beta_U$ to ensure trigger ineffectiveness, implying that all perturbed triggers may be fuzzy. For a backdoor with \metricname approaching 100\%, $\beta_U$ approaches 0. This can be interpreted as all perturbed triggers with $\|\delta\|_p > 0$ are ineffective, indicating that the backdoor only allows activation by the original trigger, making the trigger unique.

However, excessively high \metricname can lead to reduced robustness of triggers, where even minor perturbations can render triggers ineffective. Conversely, too low \metricname may result in numerous fuzzy triggers within the $\beta_U$ perturbation range of the original trigger, hindering the concealment of the backdoor. Therefore, assessing and adjusting backdoor \metricname is a key focus of our research.

\section{Security Settings}\label{sec:overview} 

In this work, we focus on the most prevalent form of backdoor attack in computer vision, specifically poisoning-based backdoor attacks \cite{gu2017badnets,chen2017targeted,nguyen2020input,li2021invisible}. The proposed \metricname metric for \backdoorname is agnostic to trigger and backdoor types, making it applicable for evaluating and enhancing most of the existing backdoor attacks. In the following, we introduce the attacker's goal, scenarios and capabilities. 

\noindent{\bf Attacker's Goal.}
The primary goal of the attacker is to embed a highly exclusive backdoor into the model and precisely define the activation conditions of the backdoor by attenuating the association of the backdoor with fuzzy triggers. This ensures that fuzzy triggers lose their ability to activate the backdoor. In essence, the backdoor in the model will only be activated when the sample carries relatively precise triggers, thereby executing the attacker's predefined malicious behaviors.

\textcolor{black}{\noindent{\bf Attack Scenarios and Capabilities.}
We consider two common scenarios for backdoor attacks: \textbf{model outsourcing} \cite{chen2017targeted, chou2020sentinet} and \textbf{data outsourcing} \cite{gu2017badnets, liu2019abs}. 
In the model outsourcing scenario, the victim may lack domain expertise and/or computational resources but has data resources. As a result, the victim may consider outsource the model training to a third party which may not be fully trusted \cite{li2022backdoor} or directly use third-party models provided by popular platforms such as HuggingFace \cite{jiang2022empirical}. In this scenario, the attacker enhances backdoors by manipulating the data and adding loss terms. For defenses, the victim does not have access to the tampered training data, but does have access to the trained model. The victim may consider implement model-level defenses based on a small portion of benign data, such as Fine-Pruning \cite{liu2018fine}, NC \cite{wang2019neural}, ABS \cite{liu2019abs}, MNTD \cite{xu2021detecting}, and MOTH \cite{tao2022model}, to mitigate the risks associated with outsourcing models.}

\textcolor{black}{In the data outsourcing scenario, the victim has computational resources and expertise to train the model but faces challenges due to insufficient training data. Consequently, the victim may consider outsource the training data preparation to a third party (e.g., Amazon Mechanical Turk \cite{turk2012amazon}) or collect data from public internet resources \cite{li2022backdoor}. In this scenario, the attacker would provide toxic data directly to the victim or spread poisoned data over the internet, but has no capability to manipulate the training process of the model. Therefore, for defenses, the victim has access to the training data and the model, but has difficulty in ensuring the dataset is clean. Therefore, the victim may consider data-level defenses to filter the poisoned samples, such as STRIP \cite{gao2019strip} and SentiNet \cite{chou2020sentinet}, and employ model-level defenses only when a small clean dataset is available.}



\section{Methodology}\label{sec:design} 

\subsection{Methodology Overview}\label{sec:overview}
\backdoorname tackles the challenge of quantifying and improving the \metricname of backdoors, with the objective of enabling backdoors to respond precisely to triggers. Our approach is centered on devising a novel metric for assessing the \metricname of backdoor attacks and introduces a corresponding attack method to enhance this characteristic.

For the measurement of \metricname, calculating the trigger upper bound for a given poisoned sample is a critical step in the algorithm. However, similar to the challenges in computing precise robust bounds \cite{katz2017reluplex}, calculating an accurate trigger upper bound faces various hurdles. Firstly, the exact computation of the trigger upper bound may involve complex mathematical forms and high-dimensional spaces, making it challenging to obtain accurate closed-form solutions through analytical methods. Secondly, the complexity and non-linear nature of deep learning models increase the computational complexity of trigger upper bound calculation, making it difficult to directly establish a mathematical model. Additionally, the calculation method for trigger upper bounds may need customization for different model architectures and datasets, adding diversity and complexity to the computation.

To address these challenges, we transform the precise calculation of the trigger upper bound into an approximate optimization problem. The primary advantage of this optimization approach is its iterative approximation of the trigger upper bound, allowing better adaptation to challenges in different scenarios, including distinct neural network structures, input data distributions, and backdoor trigger designs. This provides a more universal and flexible method, effectively evaluating and calculating trigger upper bounds.
Specifically, by formalizing the process of solving trigger upper bounds as an optimization procedure, we can formulate an objective function to search for the maximum perturbation, ensuring the trigger remains a fuzzy trigger after enduring this perturbation. In other words, the optimization objective is to find a fuzzy trigger with the maximum perturbation, implying that any trigger with a larger perturbation is distorted. Therefore, the maximum perturbation obtained through optimization can be approximately considered as a trigger upper bound.

The core idea behind \metricname enhancement is to suppress the association of the backdoor with fuzzy triggers, thereby mitigating the potential impact of most fuzzy triggers. This method aims to refine the activation conditions of the backdoor by effectively narrowing down the applicability of fuzzy triggers. To achieve this goal, two distinct sets of poisoned samples are constructed during the data poisoning process, namely, dirty samples and cover samples.
Dirty samples carry the original trigger, with their labels modified to the target labels, intending to train the model to associate the original trigger with the target category. On the other hand, cover samples carry fuzzy triggers while retaining their ground truth labels. The purpose is to decouple the backdoor's association with fuzzy triggers, thus enhancing the backdoor \metricname.

\begin{figure}[t]
    \includegraphics[trim=0 0 0 0,clip,width=0.45\textwidth]{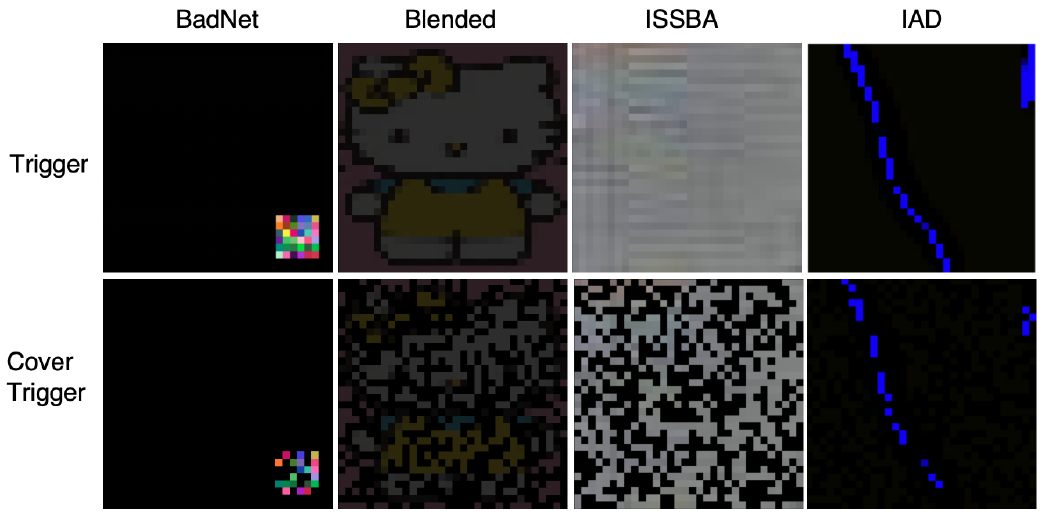}
        \caption{Demonstrations of cover triggers.}
    \label{fig:mask_trigger}
\end{figure}

It is noteworthy that the fuzzy triggers carried by cover samples are generated using a masking-based approach. In this process, a portion of the original trigger is randomly masked to create multiple distinct fuzzy triggers, with the default masking ratio set to 20\%. Figure \ref{fig:cover_trigger} illustrates an example of a fuzzy trigger in the cover sample.
Therefore, the complete poisoned training set consists of three parts: a clean dataset $(x, y) \in D$, a dirty dataset $(x_t, y_t) \in D_t$, and a cover dataset $(x_c, y) \in D_c$. In the experiments below, the poisoning rate is composed of the dirty sample rate and the cover sample rate, with the default ratio being 50:50.

\begin{figure*}
    \centering
    \includegraphics[trim=0 0 0 0,clip,width=0.9\textwidth]{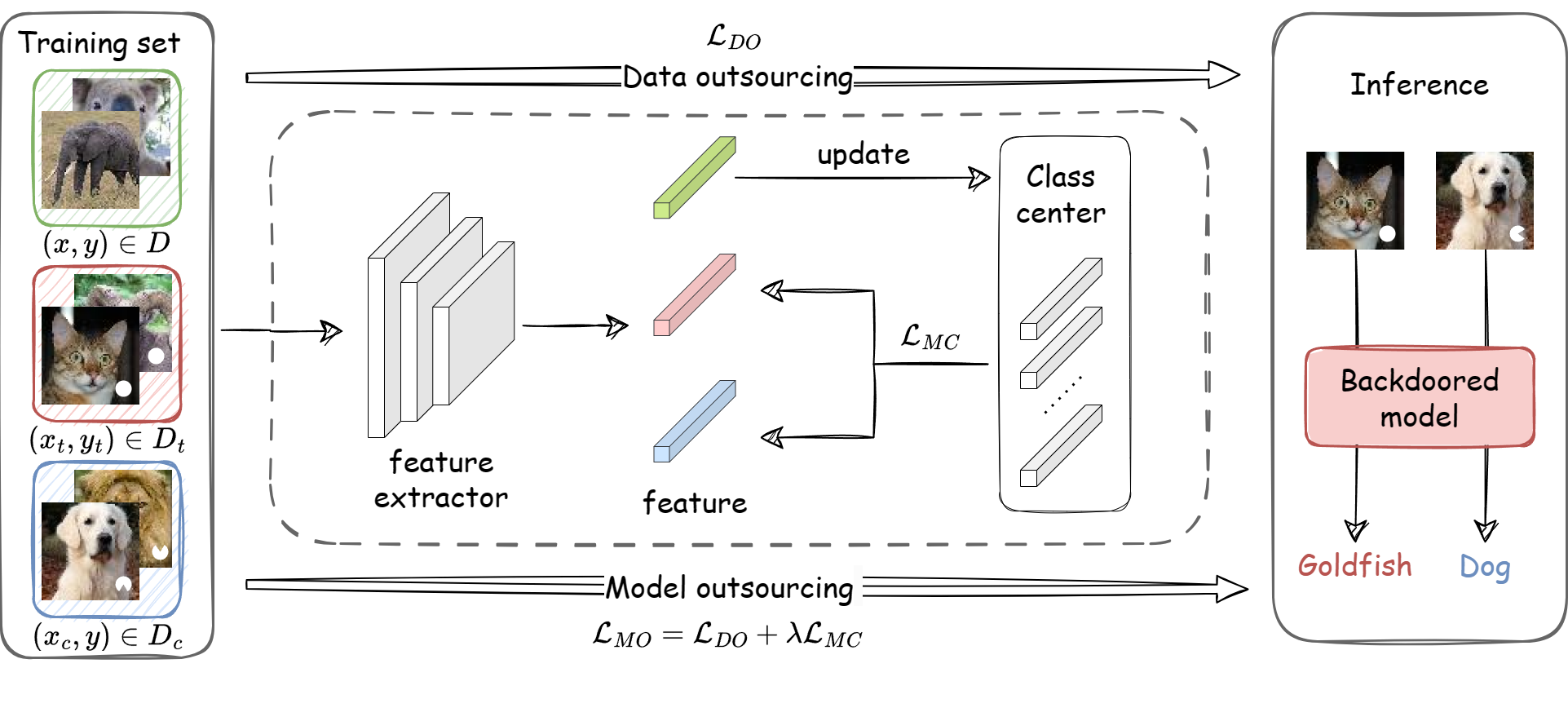}
        \caption{\textcolor{black}{Overview of \backdoorname in data outsourcing and model outsourcing scenarios.}}
    \label{fig:overview}
\end{figure*}

\noindent\textbf{Pipeline Overview.} The entire injection process of \backdoorname is illustrated in ~\autoref{fig:overview}. 
\textcolor{black}{Regular backdoor attacks \cite{chen2017targeted, gu2017badnets, li2020invisible, nguyen2020input} involve mixing poison samples ($D_t$) which carry triggers (e.g., a white circle) and incorrect labels into the training set. Beyond this, BELT also mixes cover samples ($D_c$) with fuzzy triggers (e.g., incomplete white circles) and correct labels to enhance the backdoor exclusivity. The training set comprises the clean dataset $D$, the poison dataset $D_t$, and the cover dataset $D_c$, which implant a highly exclusive backdoor in the victim model through standard training procedures.}
We consider two prevalent attack scenarios for \backdoorname: data outsourcing and model outsourcing. In the data outsourcing scenario, attackers can influence the training data but cannot manipulate the loss function. The backdoored model in the data outsourcing scenario is optimized using the loss formula \autoref{eq:DOloss}. In the model outsourcing scenario, attackers can further enhance the backdoor attack effect by utilizing the momentum center loss \autoref{eq:MCloss} to increase the distance between triggers and blurred triggers in the feature space. Finally, the backdoored model is released into public communities such as HuggingFace or GitHub, leading to unpredictable consequences \cite{jiang2022empirical}.

\subsection{Backdoor Exclusivity}
Our initial implementation of the backdoor \metricname metric seeks to employ a simple yet intuitive objective function, aiming to identify a fuzzy trigger with the maximum perturbation as a representation of the trigger upper bound for a given sample, as defined in Definition \ref{def:three}. In this context, the intuitive objective function is expressed as follows:
\begin{equation}\label{eq:init_speloss}
\min_{\|\delta\| \leq \|\delta_b\|} \left( -y_t \log(F_{\theta'}(x_{t'})) - \lambda \|\delta\|_2 \right)
\end{equation}
\begin{equation}\label{eq:disturbance_trigger}
x_{t'} = \mathbf{T'}(x, m_x, p, m_p, \delta) = x \cdot m_x + (p + \delta) \cdot m_p,
\end{equation}
where $y_t$ represents the target label, $x_{t'}$ denotes the perturbed input carrying the perturbation $\delta$, and $F_{\theta'}(x_{t'})$ signifies the predicted probability distribution by the backdoored model $F_{\theta'}$ for $x_{t'}$. Additionally, $\lambda$ serves as the balancing weight for the regularization term.

The first term in \autoref{eq:init_speloss} corresponds to the cross-entropy loss, measuring the disparity between the predicted probability distribution of the model and the true distribution. This term promotes the model's tendency to predict $x_{t'}$ as $y_t$, ensuring that the perturbed trigger remains a fuzzy trigger capable of activating the backdoor.
The second term introduces a regularization element associated with the perturbation $\delta$, where $\|\delta\|_2$ represents the $l_2$-norm of $\delta$. This term aims to control the magnitude of the perturbation, imposing a penalty proportional to the size of $\delta$. The negative sign emphasizes that smaller norms of the perturbation incur larger penalties.
In summary, the objective function comprises both cross-entropy loss and perturbation regularization terms. The former ensures the persistence of a fuzzy trigger, while the latter encourages the pursuit of a perturbation with the maximum feasible magnitude. The overarching optimization objective is to identify a perturbed trigger with the maximum perturbation. The rationale is that discovering such a fuzzy trigger with the largest perturbation implies that any further increase in perturbation would render the trigger ineffective. Therefore, the perturbation obtained upon optimization can be reasonably regarded as an approximation of the trigger upper bound.

However, this implementation did not reliably succeed in our experiments, and the optimization of this intuitive objective function often led to convergence on suboptimal solutions. 
One issue was the lack of a clear optimization direction for the regularization term, which solely considered the magnitude of the perturbation without imposing constraints on its optimization direction. Consequently, the optimization process had the potential to adjust the perturbation in arbitrary directions, possibly selecting less effective perturbation directions and resulting in the perturbation easily falling into a local optimum.
Moreover, the choice of an inappropriate value for the hyperparameter $\lambda$ could lead to suboptimal performance in balancing the cross-entropy loss and the regularization term. A larger $\lambda$ might inadequately optimize the loss function with respect to the cross-entropy term, thereby failing to ensure the persistence of the fuzzy nature of the perturbed trigger. Conversely, a smaller $\lambda$ could overlook the penalty for the perturbation, failing to effectively approximate the trigger upper bound.

To address these shortcomings, we propose two improvements to guide the perturbation in a more effective way towards approximating the trigger upper bound during the optimization process. 

\noindent{\bf Optimization direction.}
We first calculate the maximum perturbation of the trigger within a valid numerical range, referred to as the perturbation boundary $\delta_b$ (as defined in Definition \ref{def:one}). \textcolor{black}{It is worth noting that the valid numerical range is the permissible pixel value of the trigger, denoted as [$t_{min}$, $t_{max}$], which varies according to different backdoor definitions.} We use this perturbation boundary as the guiding direction for the perturbation $\delta$ during the optimization process. This enables the perturbation to be adjusted along the direction of the perturbation boundary, providing more optimization space to find a trigger with a larger perturbation.

\noindent{\bf Dynamic weight $\lambda_{dyn}$.}
Furthermore, $\lambda$ is crucial for balancing the contributions between the cross-entropy loss term and the regularization term. As defined in Definition \ref{def:three}, the trigger upper bound is considered as the one among all fuzzy triggers with the maximum perturbation. This implies that such a fuzzy trigger, with just a tiny perturbation, is enough to surpass the trigger's upper bound and become an ineffective trigger. Therefore, the sample $x_{t'}$ with a maximally perturbed fuzzy trigger should exhibit only a slightly higher probability of predicting the target label $y_t$ in its predicted probability distribution $y' = F_{\theta'}$ compared to the maximum probability of excluding $y_t$. This can be expressed as $y'_{y_t} - \underset{i \neq y_t}{\text{max}}(y'_i) < \epsilon$, where $\epsilon$ represents a very small positive number.

Based on this intuition, we design an adaptive dynamic update process for the weight $\lambda$ based on the predicted probability distribution. Specifically, in the early stages of training, we initialize $\lambda$ to a small static value (e.g., 0.1), aiming to provide a stable optimization direction for the perturbation and facilitate faster convergence of the algorithm to appropriate solutions.
As training progresses, $\lambda$ is dynamically updated based on the optimization progress and the model's feedback dynamic response $y'_{y_t} - \underset{i \neq y_t}{\text{max}}(y'_i)$. This mechanism allows $\lambda$ to adaptively update itself, providing more flexibility to adjust the impact of regularization and offering additional opportunities to escape potential local minima.

\noindent{\bf Objective function for upper bounds on triggers.}
By introducing this optimization direction and dynamic $\lambda$, we can more precisely guide the optimization process, ensuring that the increase in perturbation is primarily focused on maximizing the perturbation. This allows the perturbation of the fuzzy trigger to approach the trigger upper bound as closely as possible, thereby enhancing the algorithm's performance. 
The entire optimisation process for triggering the upper bound is described in Algorithm \ref{alg:example} in \autoref{alg}.Our final objective function is specified as follows:
\begin{equation}\label{eq:speloss}
\min_{\|\delta\| \leq \|\delta_b\|} \left( -y_t \log(F_{\theta'}(x_{t'})) + \lambda_{dyn} \|\delta_b - \delta\|_2 \right).
\end{equation}

\noindent{\bf Calculate backdoor \metricname.}
The optimization algorithm for the trigger upper bound supports our ability to optimize a perturbation to approximate the upper bound of the trigger for a single poisoned sample. Leveraging ~\autoref{eq:sample_spe}, we can utilize the optimised maximum perturbation $\delta_{max}$ to compute the \metricname of this sample. The \metricname of the backdoor is evaluated on a set containing $n$ poisoned samples (where $n$ is set to 100 by default), and the specific calculation formula for this metric is expressed as follows:
\begin{equation}\label{eq:exclusivity}
\textcolor{black}{\text{Excl} = \frac{1}{n} \sum_{i=1}^n (1 -  (\frac{\|\delta_{i, max}\|_2}{\|\delta_{i, b}\|_2})^{(t_{i, max}-t_{i, min})\sqrt{N_i}}),}
\end{equation}
\textcolor{black}{where $\|\delta_{i, max}\|$ and $\|\delta_{i, b}\|$ represent the trigger upper bound and perturbation boundary for the $i$-th sample, respectively.} The \metricname average of the $n$ poisoned samples is used to represent the \metricname of the backdoor.

\subsection{\backdoorallname}

\noindent{\bf Data Outsourcing.}
In the context of data outsourcing scenarios, attackers lack direct access or control over the model's training process. However, they can influence the data outsourcing pipeline by injecting malicious samples into the dataset to impact the model's training, aiming to implant a highly exclusive backdoor in a model.

\backdoorname can leverage common poisoning methods to implant backdoors. As previously mentioned, the training set for the model consists of three parts: a benign dataset $D$, a dirty dataset $D_t$, and a cover dataset $D_c$. In the data outsourcing scenario, the loss function $L_{DO}$ for \backdoorname can be expressed as:
\begin{equation}\label{eq:DOloss}
\begin{aligned}
\mathcal{L}_{DO} & =
\sum_{x \in D} \mathcal{L}_{ce}\left(F_{\theta'}(x), y\right) \\
& +\sum_{x_t \in D_t} \mathcal{L}_{ce}\left(F_{\theta'}\left(x_t\right), y_t\right) \\
& + \sum_{x_c \in D_c} \mathcal{L}_{ce}\left(F_{\theta'}\left(x_c\right), y\right).
\end{aligned}
\end{equation}

Here, $\mathcal{L}_{ce}$ represents the standard cross-entropy loss, $F_{\theta'}$ is the backdoored model, $x$ denotes a clean sample, $y$ is its ground truth label, $x_t$ is a dirty sample with the target label $y_t$, and $x_c$ is a cover sample with a fuzzy trigger and a correct label. The loss function is constructed by summing the cross-entropy losses over the three subsets of data. The first term ensures the model correctly classifies clean samples, encouraging the model to maintain accuracy on benign data. The second term requires the model to assign the target label to samples carrying the trigger, aiming to inject the backdoor. The third term focuses on rendering the fuzzy trigger incapable of activating the backdoor and collaborates with the second term to enforce the backdoor's connection only to the original trigger, enhancing the backdoor \metricname.

\noindent{\bf Model Outsourcing.}
In the scenario of model outsourcing, attackers have control over the complete model training pipeline. Therefore, they can not only implant a backdoor into the model following the training process outlined in the previous section for data outsourcing but can also devise additional loss terms to further enhance the performance of \backdoorname.

To this end, we propose a Momentum Center Loss, which enlarges the distance in the feature space between the original trigger and fuzzy trigger for poisoned samples (including dirty and cover samples) by converging their features towards their respective class centers. This loss term is derived from the center loss \cite{wen2016discriminative}, with two primary distinctions. Firstly, class feature centers are computed only on clean samples rather than all samples, eliminating interference from trigger features on the centers. It operates exclusively on poisoned samples, avoiding any impact on the representation of benign samples. Secondly, class centers are dynamically updated in real-time during the training process with momentum, rather than through optimization, saving computation overhead and avoiding potential local optima for class centers. The update mechanism for class centers is defined as:
\begin{equation}\label{eq:center_update}
C_i \leftarrow \text{m} \cdot C_i + (1 - \text{m}) \cdot f_i,
\end{equation}
where \(C_i\) is the center for the \(i\)-th class, \(f_i\) is the feature of a sample from class \(i\), and $\text{m}$ is the momentum factor, which is set to 0.99 by default.

The motivation behind the introduction of Momentum Center Loss is twofold. First, it promotes compact trigger features by introducing class centers in the feature space, making poisoned samples from the same class more compact and concentrated. This concentration enhances the discriminative ability of triggers in the feature space, leading to more accurate identification of the original trigger. Second, it restricts decision boundaries during training by bringing poisoned samples of the same class closer in the feature space. This limitation on decision boundaries focuses the model on recognizing subtle feature differences between the original trigger and the fuzzy trigger, making it difficult to associate fuzzy triggers with the backdoor. The objective function of Momentum Center Loss can be formulated as follows:
\begin{equation}\label{eq:MCloss}
\begin{aligned}
\mathcal{L}_{MC} & =
\sum_{x_t \in D_t} \mathcal{L}_{mse}\left(A\left(x_t\right), C_{y_t}\right) \\
& + \sum_{x_c \in D_c} \mathcal{L}_{mse}\left(A\left(x_c\right), C_{y}\right),
\end{aligned}
\end{equation}
where $\mathcal{L}_{mse}$ represents the mean square error loss, \(A(x)\) represents the feature of \(x\), \(A(\cdot)\) is the part of the network $F_{\theta'}$ used by the backdoored model to extract features, and the remaining network is denoted by \(B(\cdot)\). The forward process of the model \(F_{\theta'}(x)\) can be expressed as \(B(A(x))\). In our experiments, \(A(\cdot)\) and \(B(\cdot)\) are assumed to be divided at the last convolutional layer. The total loss function for \backdoorname in the model outsourcing scenario (\(\mathcal{L}_{MO}\)) is a combination of \(\mathcal{L}_{DO}\) and \(\mathcal{L}_{MC}\) and can be represented as:
\begin{equation}\label{eq:MOloss}
\mathcal{L}_{MO} =\mathcal{L}_{DO} + \lambda \mathcal{L}_{MC},
\end{equation}
where \(\lambda\) is the weight balancing the two losses (e.g., defaulting to \(\lambda = 1\)).

In summary, $\mathcal{L}_{MC}$ achieves improved performance by constraining poisoned samples in the feature space during training. This constraint enhances the model's ability to accurately identify the original trigger while simultaneously diminishing the association of fuzzy triggers with the backdoor, thereby improving backdoor \metricname. Furthermore, it effectively prevents poisoned samples from exhibiting a separation phenomenon from benign features in the feature space. This provides more opportunities for \backdoorname to evade countermeasures based on feature detection.

\section{Evaluation and Analysis}\label{sec:eva}

\subsection{Overview of Evaluation}

We evaluate the effectiveness of the \backdoorname on three datasets (refer to \autoref{tab:dataset}) using four representative backdoor attacks (as detailed in \autoref{sec:useb_eva}). These attacks range from input-agnostic to input-aware. We also assess the proposed \metricname metric for these backdoor attacks. Furthermore, we enhance the efficacy of \backdoorname under \textcolor{black}{seven} SOTA backdoor defense frameworks (discussed in \autoref{sec:counter}).

\noindent{\bf Datasets.}\label{subsec:setup}
We employ three standard datasets in the evaluation: CIFAR-10\cite{krizhevsky2009learning}, German Traffic Sign Recognition Dataset\cite{stallkamp2012man} (GTSRB), and TinyImageNet\cite{le2015tiny}. CIFAR-10 consists of 60,000 32x32 color images across 10 classes, serving as a benchmark for image classification tasks. GTSRB focuses on traffic sign recognition, containing over 50,000 images of 43 different traffic sign classes, making it a standard dataset for traffic sign detection and classification. TinyImagenet is a subset of the larger ImageNet dataset, comprising 200 classes with 500 training images each, offering a more manageable but still challenging dataset for image classification tasks. \autoref{tab:dataset} summarizes the statistics and the basic information of the
three datasets.

\begin{table}
\centering 
\caption{Summary of Datasets.}
\resizebox{0.50 \textwidth}{!}
{
\begin{tabular}{l   c   c   c}
\toprule 
{} & \textbf{CIFAR-10}\cite{krizhevsky2009learning}  &  \textbf{GTSRB}\cite{stallkamp2012man} & \textbf{TinyImageNet}\cite{le2015tiny}\\

\midrule
\multirow{1}*{\textbf{Task}} 
& Daily Object & Traffic Sign & General Categories \\
\multirow{1}*{\textbf{\# of Class}} 
& 10 & 43 & 200 \\
\multirow{1}*{\textbf{Target Class}} 
& \textit{automobile} & \textit{30 kph speed limit sign} & \textit{gecko} \\
\multirow{1}*{\textbf{\# of Samples}} 
& $60$K & $50$K & $110$K \\
\multirow{1}*{\textbf{Input Size}} 
& $3\times{32}\times{32}$ & $3\times{32}\times{32}$ & $3\times{64}\times{64}$ \\
\bottomrule
\end{tabular}
}
\label{tab:dataset}
\end{table}

\noindent{\bf Model Architectures.}
We conduct experiments on three popular CNN architectures for each dataset. For CIFAR-10, we utilize ResNet-18 \cite{he2016deep} for classification. For GTSRB, we opt for the VGG-11\cite{simonyan2014very} model, one of the state-of-the-art deep CNN architectures. For TinyImageNet, given the large scale and diverse categories, we leverage the EfficientNet-B3\cite{zhou2020efficient} model, a faster and smaller architecture on ImageNet scale. We train all the models from scratch.

\noindent{\bf Training Setup.}
For model training, we configure the batch size to 128 and employ the SGD optimizer with a learning rate of $1 \times 10^{-1}$ for 100 epochs of optimization. In the case of input-aware backdoor attacks, namely ISSBA and IAD, we adhere to the encoder structures specified in their respective original papers. In the assessment of backdoor defenses, we follow the settings outlined in the original papers. For optimizing the \metricname, we utilize the Adam optimizer with a learning rate set to $1 \times 10^{-3}$.
Specifically, we set the poison rate for BadNet and Blended to 0.01. For IAD and ISSBA, the poison rates are set to 0.02 and 0.1, respectively. \textcolor{black}{Cover samples are constructed based on the cover triggers, and the mask rate of triggers is set to 0.2.}

\noindent{\bf Performance Metrics.}
We measure the performance of \backdoorname following the conventional evaluation protocol for backdoor attacks and the proposed exclusivity indicator.
\begin{itemize}[leftmargin=*]
    \item \textit{Clean Dataset Accuracy (CDA):} CDA measures the percentage of clean samples that can be correctly classified. Usually, to guarantee the stealthiness of backdoor attacks, the impact on CDA should be minimal.

    \item \textit{Attack Success Rate (ASR)}: ASR measures the percentage of trigger samples misclassified into the target class by a backdoor model. Higher ASRs indicate more effective backdoor attacks.
    
    \item  \textit{Exclusivity (Excl)}:
    Excl gauges the precision of the backdoor activation conditions. A higher \metricname signifies fewer instances of potential fuzzy triggers.
    We calculated the Excl metric according to Alg.\ref{alg:example}.
\end{itemize}

\noindent{\bf Experimental Environments.}
All experiments are conducted on a server equipped with two Intel(R) Xeon(R) Silver 4210 CPU 2.20GHz 40-core processors, and six Nvidia GTX2080Ti GPUs.

\begin{table*}[]
\centering
\tabcolsep=0.0192\linewidth
\caption{\textcolor{black}{Comparison of four existing backdoor attacks before and after BELT  {\footnotesize ($^+/^{++}$ denotes data/model outsourcing scenarios)}. }}
\label{tab_main}
\begin{tabular*}{\linewidth}{@{}llllllllll@{}}
\toprule
\multirow{2}{*}{\textbf{Backdoor Attack Type}} & \multicolumn{3}{c}{CIFAR-10-ResNet18} & \multicolumn{3}{c}{GTSRB-VGG11} & \multicolumn{3}{c}{TinyImageNet-EN-B3} \\ \cmidrule(r){2-4}  \cmidrule(r){5-7} \cmidrule(r){8-10}
{} & CDA(\%) & ASR(\%) & Excl(\%) & CDA(\%) & ASR(\%) & Excl(\%) & CDA(\%) & ASR(\%) & Excl(\%) \\ \cmidrule(r){1-1} \cmidrule(r){2-4} \cmidrule(r){5-7} \cmidrule(r){8-10}

{Benign} & 94.69 & - & - & 97.81 & - & - & 53.90 & - & -  \\ 
\cmidrule(r){1-1} \cmidrule(r){2-4}  \cmidrule(r){5-7} \cmidrule(r){8-10}

{BadNet\cite{gu2017badnets}} & 93.17 & \textit{100.00} & 0.00 & 97.12 & 99.15 & 0.00 & 50.95 & \textit{100.00} & 34.01  \\ 
{BadNet$^+$} & \textit{94.85} & 99.98 & 84.55 & 98.12 & 99.99 & 84.14 & \textit{52.79} & \textit{99.99} & 67.19 \\ 
{BadNet$^{++}$} & 94.62 & \textit{100.00} & \underline{\textbf{87.09}} & \textit{98.35} & 99.98 & \underline{\textbf{89.34}} & 52.31 & \textit{100.00 }& \underline{\textbf{72.30}} \\  \cmidrule(r){1-1} \cmidrule(r){2-4}  \cmidrule(r){5-7} \cmidrule(r){8-10}

{Blended\cite{chen2017targeted}} & \textit{95.28} & \textit{99.79} & 0.99 & \textit{98.37} & 99.27 & 3.92 & 51.77 & \textit{99.58} & 29.27  \\ 
{Blended$^+$} & 94.48 & 98.47 & 16.18 & 96.41 & 99.37 & 20.70 & \textit{51.95}  & 98.78 & 49.22  \\ 
{Blended$^{++}$} & 94.61 & 99.14 & \underline{\textbf{20.06}} & 97.70 & \textit{99.98} & \underline{\textbf{24.21}} & 51.31  & 99.29 & \underline{\textbf{53.98}} \\  \cmidrule(r){1-1} \cmidrule(r){2-4}  \cmidrule(r){5-7} \cmidrule(r){8-10}

{ISSBA\cite{li2021invisible}} & 94.80 & 99.03 & 21.98 & 97.64 & 99.89 & 23.34 & 48.70 & \textit{100.00} &  43.00 \\ 
{ISSBA$^+$} & \textit{95.16} & 97.57 & 23.73 & \textit{98.12} & 99.97 & 25.76 & \textit{48.86} & 99.91 & 46.66  \\ 
{ISSBA$^{++}$} & 94.13 & \textit{99.97} & \underline{\textbf{27.33}} & 98.01 & \textit{100.00} & \underline{\textbf{29.68}} & 48.26 & 99.87 & \underline{\textbf{53.03}}   \\  \cmidrule(r){1-1} \cmidrule(r){2-4}  \cmidrule(r){5-7} \cmidrule(r){8-10}

{IAD\cite{nguyen2020input}} & 93.32 & \textit{99.93} & 6.18 & 97.19 & 99.25 & 14.84 & \textit{50.43} & 99.18 & 38.19  \\ 
{IAD$^+$} & 92.78 & 97.04 & 29.67 & 97.33 & 99.51 & 48.89  & 47.95 & \textit{99.47} & 55.19  \\ 
{IAD$^{++}$} & \textit{93.34} & 99.56 & \underline{\textbf{36.71}} & \textit{97.74} & \textit{99.85} &  \underline{\textbf{75.07}} & 48.93 & 98.95 & \underline{\textbf{89.26}}   \\  

 \bottomrule
\end{tabular*}
\end{table*}

\subsection{\backdoorname Evaluation and Analysis}\label{sec:useb_eva}

We evaluate the \backdoorname framework in two scenarios, i.e., data outsourcing and model outsourcing, by combining \backdoorname with four SOTA attacks. In the data outsourcing scenario, we introduce dirty and cover samples through data poisoning alone, denoted as $^+$. In the model outsourcing scenario, where the attacker has complete control over the entire model training process, we enhance the \backdoorname with momentum-centered loss (\autoref{eq:MOloss}), denoted as $^{++}$. The summarized experimental results are presented in \autoref{tab_main}.

\noindent \textbf{Benign Model.}
The results obtained from benign datasets serve as a basis for comparisons. Ideally, all backdoor attacks should exert minimal influence on CDA to ensure the usability and stealthiness of the backdoored models.

\noindent \textbf{BadNet.}
To evaluate BadNet\cite{gu2017badnets}, we employ $6 \times 6$ random blocks of pixels as triggers, as depicted in the first column of \autoref{fig:mask_trigger}. Utilizing \backdoorname, we strengthen BadNet's \metricname under data outsourcing and attack outsourcing scenarios, denoted as BadNet$^+$ and BadNet$^{++}$, respectively. From Table \ref{tab_main}, it is evident that both BadNet before and after reinforcement exhibit high ASR while maintaining a comparable CDA to the clean model.
However, analyzing the backdoor \metricname reveals a significant distinction. The regular BadNet's Excl is notably low, approaching 0\%, indicating loose activation conditions and numerous fuzzy triggers that may activate the backdoor. 
In contrast, the \metricname of BadNet$^+$ and BadNet$^{++}$ shows substantial improvement compared to regular BadNet, increasing from \textcolor{black}{0\% to 84\%-87\%}. This enhancement implies that BadNet$^+$ and BadNet$^{++}$ respond more accurately to triggers, reducing the occurrence of potentially fuzzy triggers.
We elucidate the reasons behind the regular BadNet's ease of detection from the \metricname perspective and demonstrate that \metricname-enhanced BadNet successfully evades SOTA backdoor defenses, as detailed in \autoref{sec:counter}.

\noindent \textbf{Blended.}
To evaluate Blended\cite{chen2017targeted}, we employ the \textit{hello\_kitty} image with transparency to 0.2 as triggers, as depicted in the second column of \autoref{fig:mask_trigger}. Similar to BadNet evaluation, we conduct experiments under data outsourcing and model outsourcing scenarios, denoted as Blended$^+$ and Blended$^{++}$. As shown in \autoref{tab_main}, \backdoorname demonstrates negligible effects on CDA and ASR, while significantly improving the backdoor \metricname, increasing from \textcolor{black}{0.99\% to 16\%-20\%}. It is noteworthy that the \metricname of Blended is relatively smaller than BadNet, attributed to the broader range of perturbations and fixed perturbation value of Blended trigger, the trigger itself has lower \metricname compared to BadNet, resulting in numerous potential fuzzy triggers. However, \backdoorname still significantly increases the \metricname of Blended by approximately \textcolor{black}{2-20 times} and manifests in corresponding evasion effects in the MOTH defense (\autoref{sec:counter}).

\noindent \textbf{ISSBA.}
ISSBA aims to propose an imperceptible backdoor trigger\cite{li2020invisible}. We employ a DNN-based image steganography network to generate sample-specific invisible additive noises as backdoor triggers by encoding an attacker-specified string into benign images, as shown in the third column of \autoref{fig:mask_trigger}.
According to our experimental results, ISSBA$^+$ and ISSBA$^{++}$ preserve ASR and have minimal impact on CDA. The \metricname of ISSBA$^+$ and ISSBA$^{++}$ still outweighs ISSBA. It is noted that the Excls of \textcolor{black}{regular ISSBA} are significantly greater than other attacks. This phenomenon can be attributed to two folds. Firstly, ISSBA adds very small perturbation to the images, (about -0.05 to 0.05), making them imperceptible to human eyes. This significantly reduces the range for perturbing the trigger, resulting in fewer potential fuzzy triggers and inherently higher exclusivity compared to traditional visible perturbations. Secondly, ISSBA is a sample-specific backdoor attack. 
Intuitively, sample-specific attacks create sample-specific triggers, which form a corresponding input-dependent trigger space.
This further limits the range in which the trigger can be perturbed, thereby enhancing \metricname. 

\noindent \textbf{IAD.}
To evaluate IAD\cite{nguyen2019diot}, we train a 7-layer CNN-based generator for three datasets from scratch to generate input-aware triggers for dirty samples. As documented in \autoref{tab_main}, CDA and ASR of IAD$^+$ and IAD$^{++}$ remain largely unchanged compared to IAD. The \metricname of IAD$^+$ and IAD$^{++}$ show improvement compared to IAD. In contrast to ISSBA, IAD adds more pronounced perturbations to images. However, as analyzed in ISSBA, input-aware triggers possess higher exclusivity naturally compared to Blended. Therefore, IAD$^+$ and IAD$^{++}$ possess Excls between \textcolor{black}{29\% and 89\%}. 

\subsection{Backdoor Countermeasures Evaluation}\label{sec:counter}

During the evaluation phase of countermeasures, our focus is primarily on evaluating the performance of \backdoorname based on Badnets. Badnets employ a trigger that is the simplest form of static patching, and we choose patch triggers as the focal point of evaluation for two key reasons.
Firstly, the trigger used by Badnets is a straightforward backdoor attack method, employing the simplest form of patching without relying on complex or input-aware triggers. This simplicity makes Badnets one of the few practical and effective real-world backdoor attacks~\cite{li2022backdoor,li2021backdoor}. Therefore, the contribution of \backdoorname to enhancing the stealthiness against Badnets holds practical significance.
Secondly, although countermeasures against backdoors may have varying applicability to different types of backdoor attacks depending on different threat models, many countermeasures are relatively effective against Badnets~\cite{gao2019strip,liu2019abs,wang2019neural,chou2020sentinet,tao2022model, tao2022model}. By evaluating \backdoorname based on Badnets, we can more clearly elucidate the effectiveness of \backdoorname in enhancing backdoor evasibility.
The methodology of the evaluated defense countermeasures is described in~\autoref{methodology_description}.

\subsubsection{Neural Cleanse}

\begin{figure}
    \centering
    \includegraphics[trim=0 0 0 0,clip,width=0.45\textwidth]{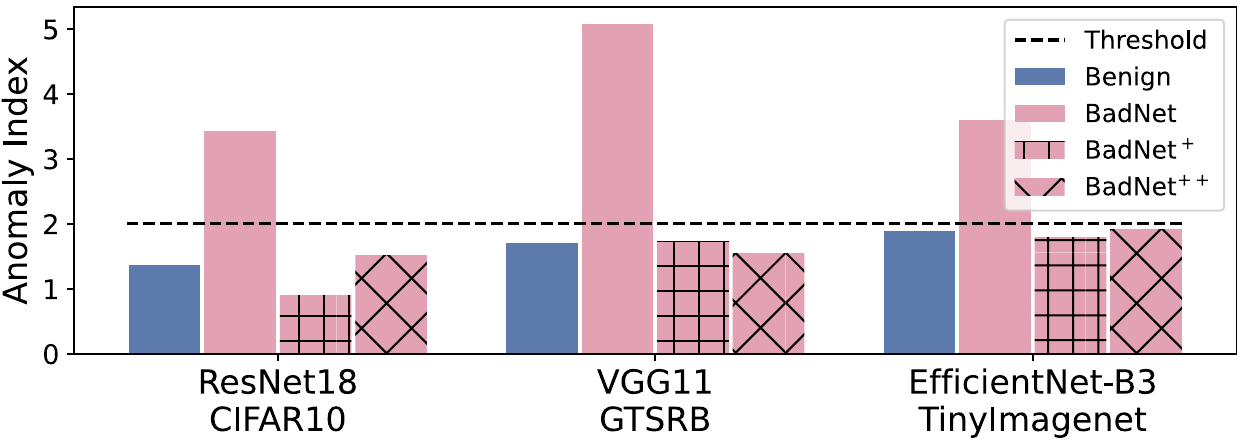}
        \caption{Performance of Neural Cleanse on backdoored models before and after \backdoorname.}
    \label{fig:NC_AI}
\end{figure}

\begin{figure}
    \centering
    \includegraphics[trim=0 0 0 0,clip,width=0.45\textwidth]{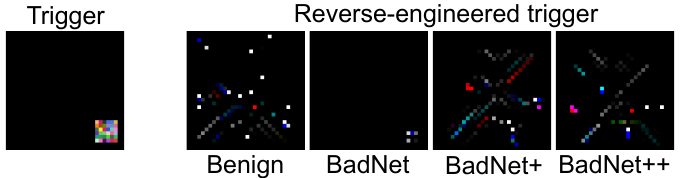}
        \caption{The trigger reverse-engineered by Neural Cleanse on the target label.}
    \label{fig:NCTrigger}
\end{figure}

We evaluate all BadNet models trained in \autoref{sec:useb_eva} using Neural Cleanse \cite{NC_git}. We conduct 100 optimization periods through a reverse engineering process to generate reverse triggers for each class of the evaluated models. The neural cleanup process is repeated 10 times for each model to obtain an average anomaly metric, as illustrated in ~\autoref{fig:NC_AI}.
The results reveal that the anomaly index of the \backdoorname model for all settings remains below the threshold of 2. This suggests that Neural Cleanse fails to detect the backdoor in the \backdoorname model, whether it is a BadNet$^+$ or a BadNet$^{++}$. The reason for this failure lies in Neural Cleanse's inability to accurately reverse engineer the fuzzy trigger through the reverse engineering process. ~\autoref{fig:NCTrigger} provides an example set of reverse triggers for the target tag on the CIFAR-10 task, where the first column represents the original trigger for the backdoor. The benign model, lacking a backdoor, presents the reverse triggers as natural features of the target class, while the BadNet model, with a \metricname of 0\%, allows fuzzy triggers to be easily reverse-engineered.
However, for BadNet$^+$ and BadNet$^{++}$, the reverse triggers of the target class closely resemble those of the benign model. This is attributed to \backdoorname significantly reducing the presence of fuzzy triggers by strengthening the backdoor \metricname, making it challenging for Neural Cleanse to identify fuzzy triggers within the narrowed exclusivity boundary during the reverse engineering process.

\subsubsection{ABS}

\begin{figure}
    \centering
    \includegraphics[trim=0 0 0 0,clip,width=0.45\textwidth]{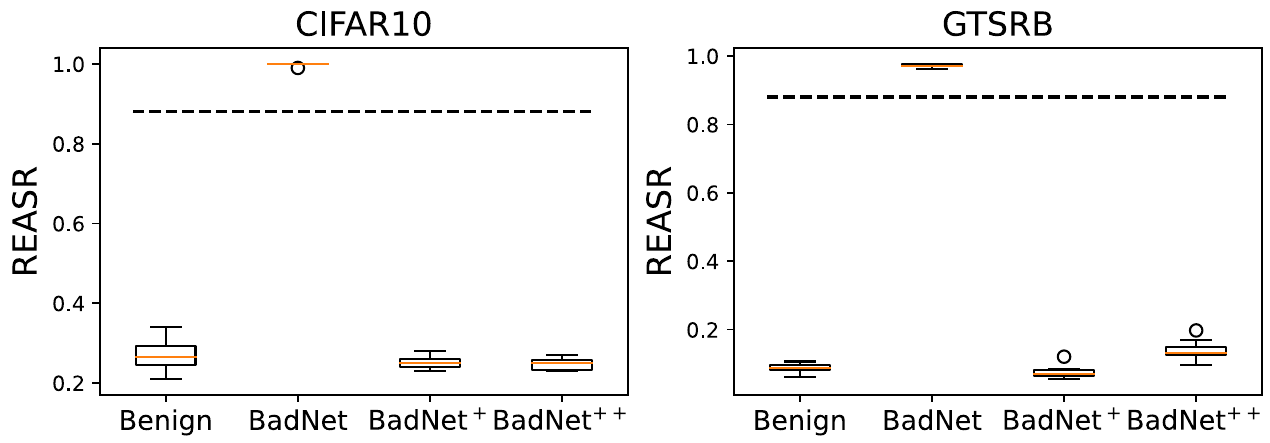}
        \caption{Performance of ABS on backdoored models before and after \backdoorname.}
    \label{fig:reasr_box}
\end{figure}

We use ABS \cite{ABS_git} to evaluate the Badnet models in \autoref{sec:useb_eva} based on training on two tasks, CIFAR-10 and GTSRB. We construct a small clean dataset by randomly sampling 10 samples from each class to support ABS execution. In total, we sample 100 samples for CIFAR-10 and 430 samples for GTSRB. Subsequently, following ABS's default settings, we detect the 10 most likely compromised neurons and reverse-engineer a corresponding trigger for each of these neurons to calculate the REASR metric. ~\autoref{fig:reasr_box} presents the REASR distribution for the 10 most likely compromised neurons in each model.

It is observed that ABS can effortlessly detect the backdoor in the regular Badnet model, as its REASR exceeds the threshold of 0.88 and approaches 1.0. However, ABS fails to effectively detect Badnet$^+$ and Badnet$^{++}$, where their REASR values are close to those of the benign model. This is not surprising; similar to Neural Cleanse, ABS performs reverse engineering on compromised neurons to obtain corresponding reverse triggers. The strong \metricname of the backdoor hinders this process, preventing it from acquiring a reverse trigger capable of activating the backdoor, resulting in REASR values far below the threshold.

\subsubsection{MNTD}

\begin{table}
\centering 
\caption{Performance of MNTD on backdoored models before and after \backdoorname.}
\resizebox{0.45 \textwidth}{!}
{
\begin{tabular}{l c c c c} %
\toprule 
Test model set & CDA(\%) & ASR(\%) & Score & AUC(\%) \\

\cmidrule(r){1-1} \cmidrule(r){2-5} 
Benign & 61.36 & - & 3.46 & \multirow{2} * {95.63}\\
Backdoored & 61.29 & 99.73 & 12.14\\

\cmidrule(r){1-1} \cmidrule(r){2-5} 
Benign & 61.36 & - & 3.46 & \multirow{2} * {75.81}\\
Backdoored$^+$ & 62.18 & 85.27 & 7.34 \\

\cmidrule(r){1-1} \cmidrule(r){2-5} 
Benign & 61.36 & - & 3.46 & \multirow{2} * {77.61}\\
Backdoored$^{++}$ & 62.44 & 95.44 & 7.55 &  \\

\bottomrule
\end{tabular}
}
\label{tab:mntd}
\end{table}

For the evaluation of Meta-Neural Trojan Detection (MNTD) \cite{MNTD_git}, we adhere to the training settings of MNTD initially, employing transfer learning to construct a meta-classifier. This meta-classifier is trained on the CIFAR-10 dataset, comprising 2048 benign models and 2048 backdoor models within a shadow model set. During the evaluation phase, we train 256 benign test models (referred to as "Benign") and 256 backdoor test models (referred to as "Backdoored"). Additionally, we train an additional 256 backdoor models in the data outsourcing scenario (referred to as "Backdoored$^+$") and another 256 backdoor models in the model outsourcing scenario (referred to as "Backdoored$^{++}$").

It is crucial to highlight that both "Backdoored$^+$" and "Backdoored$^{++}$" maintain identical training settings as "Backdoored," encompassing model architecture, backdoor configurations, and training hyperparameters. This ensures that any impact on detection performance stems solely from the actions of the backdoor itself, rather than changes in hyperparameters \cite{qiu2023towards}.
In \autoref{tab:mntd}, the average performance of the three test model sets is presented. MNTD demonstrates commendable performance on the test set containing ordinary backdoored models ("Backdoored"), achieving an AUC of 95.63\%. However, the high-\metricname backdoor models effectively resist MNTD detection, exhibiting AUCs of 75.81\% and 77.61\% for the test sets based on "Backdoored$^+$" and "Backdoored$^{++}$," respectively. The primary cause lies in the fact that backdoors with stronger \metricname exhibit reduced responsiveness to optimized queries compared to those with weaker \metricname. Consequently, meta-classifiers are inclined to predict lower scores for strongly exclusive backdoor models, posing challenges in distinguishing them significantly from benign models and leading to diminished detection performance.

\subsubsection{MOTH}
According to the original configurations outlined in MOTH \cite{MOTH_git}, we conduct experiments on both backdoored models and augmented backdoored models using three standard datasets. Following the warm-up and two-sided backdoor generation training pipeline, we adopt the pair scheduling method from the same source. Promising pairs are determined based on the largest growth of class distance within a fixed number of optimization cycles.
MOTH is evaluated on backdoored models obtained from previous experiments, and the effectiveness of our proposed \backdoorname framework is demonstrated in ~\autoref{tab_moth}. Our focus lies on examining the CDA and ASR of both original backdoored models and \backdoorname-augmented models.
The defense effects under identical experimental conditions are documented, showing the remaining CDA and ASR of the model after applying the MOTH defense. In essence, \backdoorname reinforces the backdoor \metricname, rendering the backdoor samples more stealthy. This reinforcement could potentially weaken defenses that rely on hardening the model's decision boundaries. Results in \autoref{tab_moth} reveal a significant reduction in ASR for non-augmented models following the MOTH defense. Conversely, models augmented with our \backdoorname framework exhibit a substantial increase in ASR compared to models without \backdoorname. These experimental outcomes underscore the efficacy of our proposed \backdoorname framework.

\begin{table}[]
\centering
\tabcolsep=0.020\linewidth
\caption{Performance of MOTH on backdoored models before and after \backdoorname.}
\label{tab_moth}
\renewcommand{\arraystretch}{1.2}
\begin{tabular}{@{}lllll@{}}
\toprule
\multirow{3}{*}{\begin{tabular}{@{}c@{}} Attack Type \end{tabular}} & \multicolumn{2}{c}{CIFAR-10-ResNet18} & \multicolumn{2}{c}{GTSRB-VGG11}  \\ \cmidrule(r){2-3} \cmidrule(r){4-5} 
{} & \begin{tabular}{@{}c@{}} CDA(\%) \end{tabular} & \begin{tabular}{@{}c@{}} ASR(\%) \end{tabular} & \begin{tabular}{@{}c@{}} CDA(\%) \end{tabular} & \begin{tabular}{@{}c@{}} ASR(\%) \end{tabular} \\ \cmidrule(r){1-1} \cmidrule(r){2-3} \cmidrule(r){4-5} 


{BadNet} & 92.32 & 32.47  & \textit{75.31} & 38.29 \\ 
{BadNet$^+$} & \textit{93.41} & 98.62  & 74.88 & 77.56 \\ 
{BadNet$^{++}$} & 93.03 & \underline{\textbf{100.00}} & 67.76 & \underline{\textbf{99.71}} \\ \cmidrule(r){1-1} \cmidrule(r){2-3} \cmidrule(r){4-5} 

{Blended} & \textit{93.64} & 33.74 & 76.39 & 1.30   \\ 
{Blended$^+$} & 92.34 & 96.43 & 62.61 & \underline{\textbf{98.86}}   \\ 
{Blended$^{++}$} & 91.98 & \underline{\textbf{99.53}}  & \textit{88.66} & 92.54 \\  \cmidrule(r){1-1} \cmidrule(r){2-3} \cmidrule(r){4-5} 

{ISSBA} & 81.29 & 1.20 & \textit{95.36} & 66.91  \\ 
{ISSBA$^+$} & \textit{85.61} & 87.78  & 69.86 & 74.50  \\ 
{ISSBA$^{++}$} & 84.42 & \underline{\textbf{99.05}} & 92.21 & \underline{\textbf{93.66}} \\ 
\cmidrule(r){1-1} \cmidrule(r){2-3} \cmidrule(r){4-5} 

{IAD} & \textit{92.21} & 9.78  & \textit{69.53} & 57.81    \\ 
{IAD$^+$} & 90.38 & 92.46 & 66.75 & \underline{\textbf{77.30}}  \\ 
{IAD$^{++}$} & 91.79 & \underline{\textbf{97.89}}  & 69.30  & 77.21 \\  

 \bottomrule
\end{tabular}
\end{table}

\subsubsection{STRIP}
\begin{figure}
    \centering
    \includegraphics[trim=0 0 0 0,clip,width=0.45\textwidth]{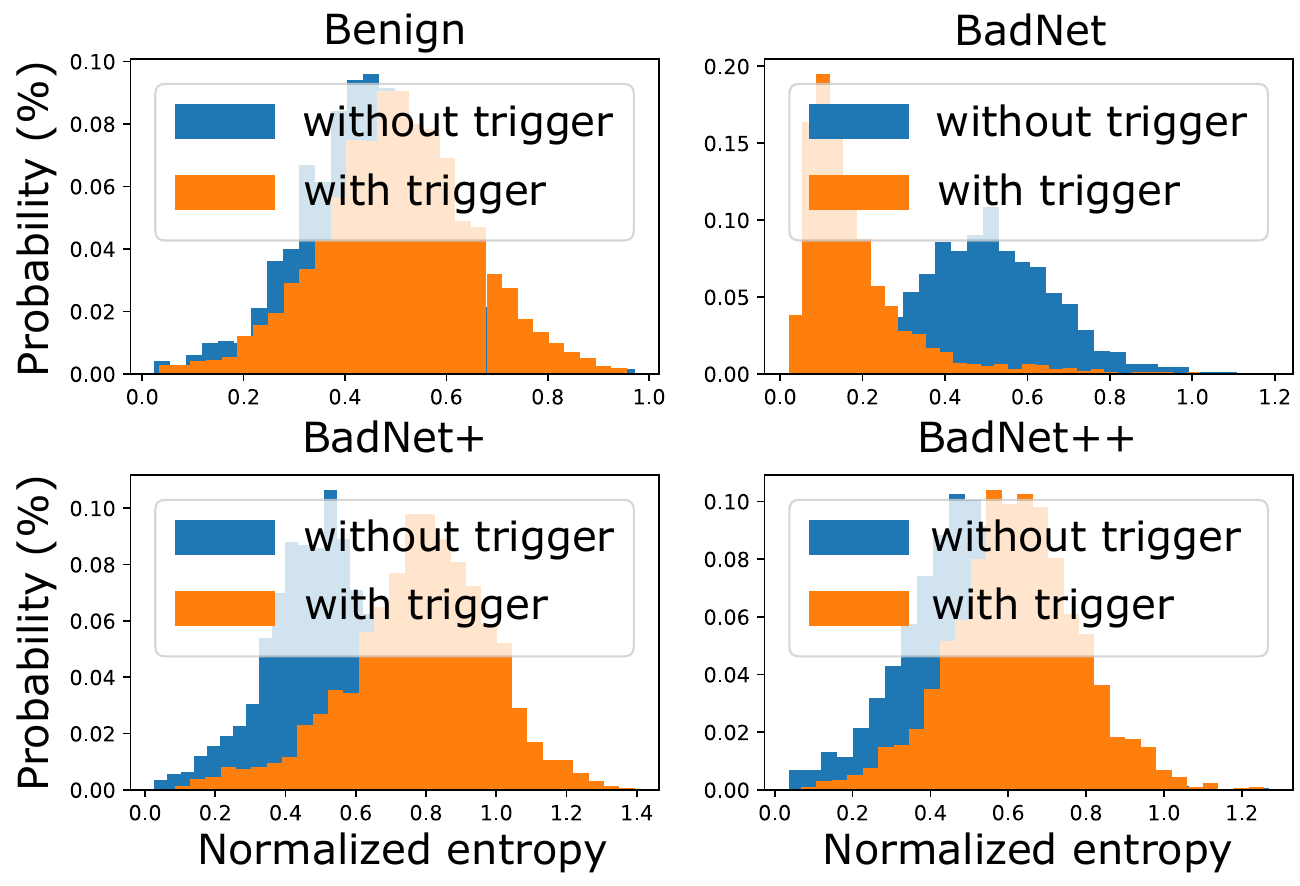}
        \caption{Performance of STRIP on backdoored models before and after \backdoorname.}
    \label{fig:strip}
\end{figure}

\autoref{fig:strip} displays the entropy distribution results of benign samples and trigger-carrying samples on a regular Badnet and a \metricname-enhanced Badnet as evaluated by STRIP \cite{STRIP_git}. The assessed models utilize the ResNet18 architecture and are trained on the CIFAR-10 dataset. The distribution is evaluated on 1000 trigger images and 1000 clean images. For the regular Badnet model, STRIP effectively distinguishes samples carrying triggers from clean samples because there is a noticeable gap in the entropy value distribution between trigger samples and clean samples. However, for Badnet$^+$ and Badnet$^{++}$, their entropy distributions are similar to the benign model, with a significant overlap, rendering STRIP ineffective.
The primary reason for this phenomenon is that the high \metricname of the backdoor disrupts the strong hijacking effect of the trigger under different perturbations. As a result, the perturbed trigger, when overlaid with benign samples, is no longer a fuzzy trigger capable of activating the backdoor.

\subsubsection{SentiNet}

\begin{figure}
    \centering
    \includegraphics[trim=0 0 0 0,clip,width=0.45\textwidth]{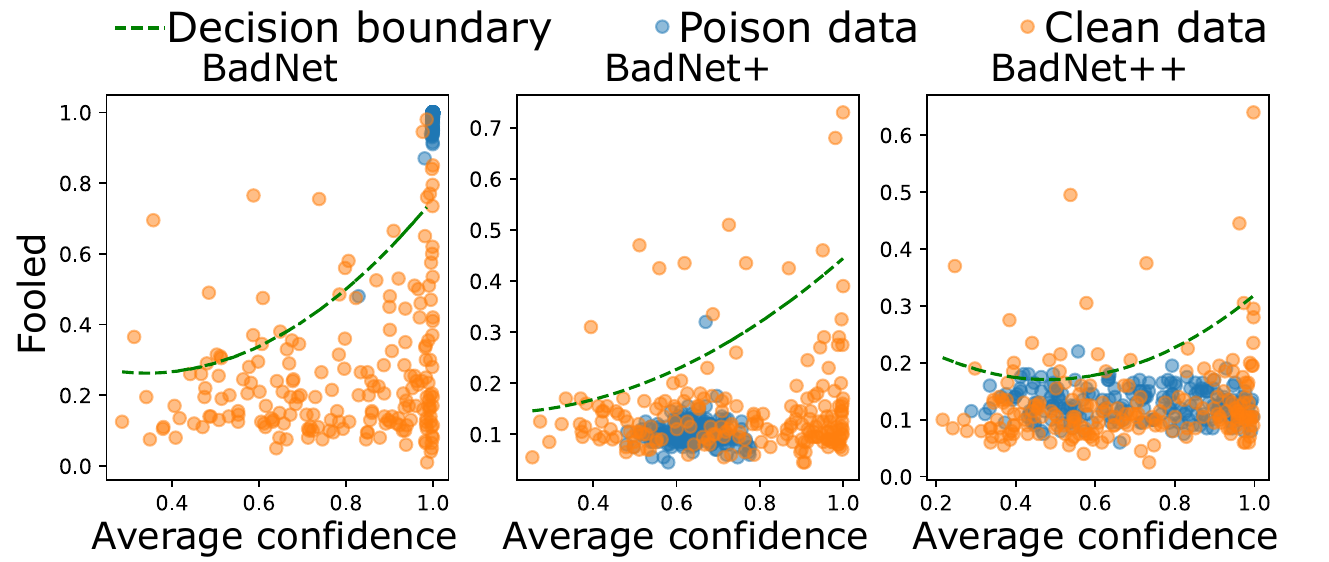}
        \caption{Performance of SentiNet on backdoored models before and after \backdoorname.}
    \label{fig:SentiNet}
\end{figure}

We partition the clean data into intervals based on a fixed average confidence step and select the two data points with the highest Foolde within each interval as decision points for fitting the decision boundary. Figure \autoref{fig:SentiNet} illustrates the detection performance of SentiNet \cite{SentiNet_git}. For a regular Badnet, the trigger on poisoned data significantly influences the model output, resulting in high average confidence and Foolde within each interval. This leads to an effective decision boundary that distinguishes poisoned data from clean data. However, the detection performance of SentiNet is not satisfactory for Badnet$^+$ and Badnet$^{++}$. Poisoned data and clean data are confounded, causing the decision boundary to be ineffective in distinguishing them. One possible reason is that GradCAM cannot ensure the capture of a complete trigger (see the last column of Figure \autoref{fig:defense_trigger}). For highly exclusive backdoors, such incomplete triggers do not elicit a response from the backdoor, escaping detection by SentiNet.

\textcolor{black}{\subsubsection{Fine-Pruning}
Fine-Pruning \cite{liu2018fine} is one of the earliest defenses based on parameter updating, which is motivated from the following observation: the inputs with triggers usually activate neurons that are dormant in the presence of clean inputs. Therefore, Fine-Pruning attempts to erase the backdoor by pruning neurons which are scarcely activated for the clean inputs. To reduce the impact on CDA, Fine-Pruning further applies fine-tuning after pruning to restore the model performance.}

\begin{figure}[t]
    \centering
    \includegraphics[trim=0 0 0 0,clip,width=0.45\textwidth]{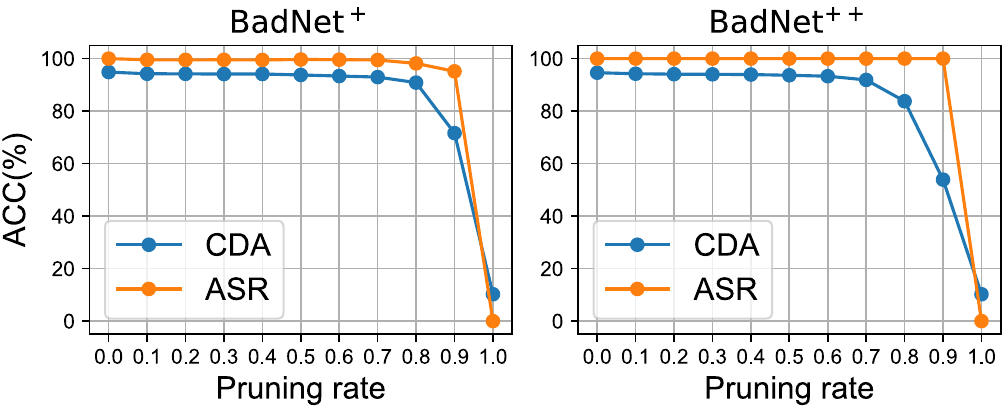}
        \caption{\textcolor{black}{Performance of Fine-Pruning on backdoored models after BELT.}}
    \label{fig:finepruning}
\end{figure}

\textcolor{black}{To verify the impact of updating model parameters on BELT, we apply Fine-Pruning (with the default configuration in its official implementation \cite{FinePruning_git}) on BadNet$^+$ and BadNet$^{++}$, with the CIFAR10 dataset and the ResNet18 architecture. As shown in \autoref{fig:finepruning}, Fine-Pruning fails to effectively remove the backdoor enhanced by BELT, since ASR remains higher than CDA at various pruning rates. Only when the pruning rate reaches 90\% does ASR begin to decrease, while CDA has already suffered an unacceptable loss.}

\section{Discussion}\label{sec:dis}

\begin{table}
\centering 
\caption{Ablation study results for trigger upper bound optimization algorithm.}
\resizebox{0.45 \textwidth}{!}
{
\begin{tabular}{l c c c} %
\toprule
\multirow{2}*{Baseline} & \multicolumn{3}{c}{Trigger upper bound $\beta_U$} \\
\cmidrule(r){2-4}
& Badnet & Badnet$^+$ & Badnet$^{++}$ \\

\cmidrule(r){1-1} \cmidrule(r){2-4} 
Comparison Group A &  5.13 & 6.19 & 6.09 \\

Comparison Group B  & 5.13 & 6.41 & 6.12 \\

Comparison Group C & 7.76 & 6.39 & 6.21 \\

\cmidrule(r){1-1} \cmidrule(r){2-4} 
\textbf{Overall} & \textbf{7.76} & \textbf{6.68} & \textbf{6.36} \\

\bottomrule
\end{tabular}
}
\label{tab:abl}
\end{table}

\noindent\textbf{(a) Ablation Studies.} We conducted the following ablation experiments to validate the effectiveness and contribution of each component in the trigger upper bound optimization process. The experimental subject is the Badnet model, with a ResNet18 architecture, trained on the CIFAR-10 dataset. The specific results are presented in \autoref{tab:abl}.

\begin{itemize}[leftmargin=*]
\item\textbf{Comparison Group A: Intuitive objective function.} First, we investigated the impact of an intuitive objective function on the optimization of the trigger upper bound. In this baseline, the optimization process does not include the optimization direction and dynamic weights. The objective function used is as defined in~\autoref{eq:init_speloss}. However, such an optimization process is prone to local optima, resulting in a trigger upper bound significantly smaller than its actual precise value, especially for a regular Badnet.

\item\textbf{Comparison Group B: Without optimization direction.} We reveal the influence of the optimization direction on the trigger upper bound. Here, we use the regularization term from~\autoref{eq:init_speloss} instead of the one from~\autoref{eq:exclusivity}, while retaining the dynamic weight mechanism, to eliminate the impact of the optimization direction on the optimization results. The results show a noticeable difference from the overall performance, indicating that the optimization direction guides the process, aiding in escaping local optima and finding a more significant perturbation for a fuzzy trigger.

\item\textbf{Comparison Group C: Without dynamic weight.} We use this baseline to validate the effectiveness of dynamic weights. Here, we remove the dynamic weight mechanism from~\autoref{eq:exclusivity} and replace it with a small static value (0.1). The table shows a significant performance drop on both BadNet$^+$ and BadNet$^{++}$, confirming that dynamic weights help bring the perturbation closer to the trigger upper bound during the optimization process.
\end{itemize}

These three baselines illustrate that the optimization direction and dynamic weight mechanism can play a positive role in the optimization of the trigger upper bound. By introducing these two components, we can effectively calculate a more accurate trigger upper bound.

\begin{figure}
    \centering
    \includegraphics[trim=0 0 0 0,clip,width=0.4\textwidth]{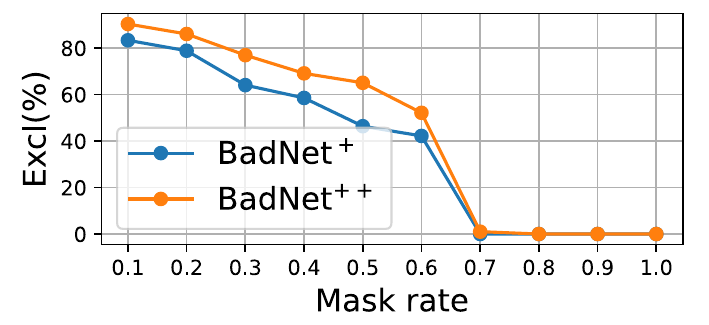}
        \caption{Comparison of backdoor \metricname of Badnet under different mask rates.}
    \label{fig:spec_mr}
\end{figure}

\noindent\textbf{(b) Effect of Mask Rate on Exclusivity}. Cover samples are instances with a fuzzy trigger and correct labels, designed to suppress the association between the backdoor and the fuzzy trigger, thereby enhancing the \metricname of the backdoor. In our approach, the trigger carried by cover samples is constructed through a masking-based method. It involves randomly masking a fixed percentage (e.g. 20\%) of the original trigger, generating different fuzzy triggers.

The quality of the fuzzy trigger carried by cover samples directly influences the degree of \metricname enhancement. Therefore, we focus on the impact of mask rates on backdoor \metricname. We trained multiple Badnet models using different mask rates, employing the ResNet18 architecture and the CIFAR-10 dataset. As depicted in~\autoref{fig:spec_mr}, a mask rate of 1.0 indicates that the fuzzy trigger on cover samples is completely masked, treating these samples as clean samples, corresponding to the training setting of a conventional backdoor. As the mask rate decreases, backdoor \metricname gradually increases. This is attributed to lower mask rates improving the quality of the fuzzy trigger, making it closer to the original trigger. By suppressing the association between the backdoor and high-quality fuzzy triggers, backdoor \metricname can be significantly enhanced.

\begin{figure}
    \centering
    \includegraphics[trim=0 0 0 0,clip,width=0.4\textwidth]{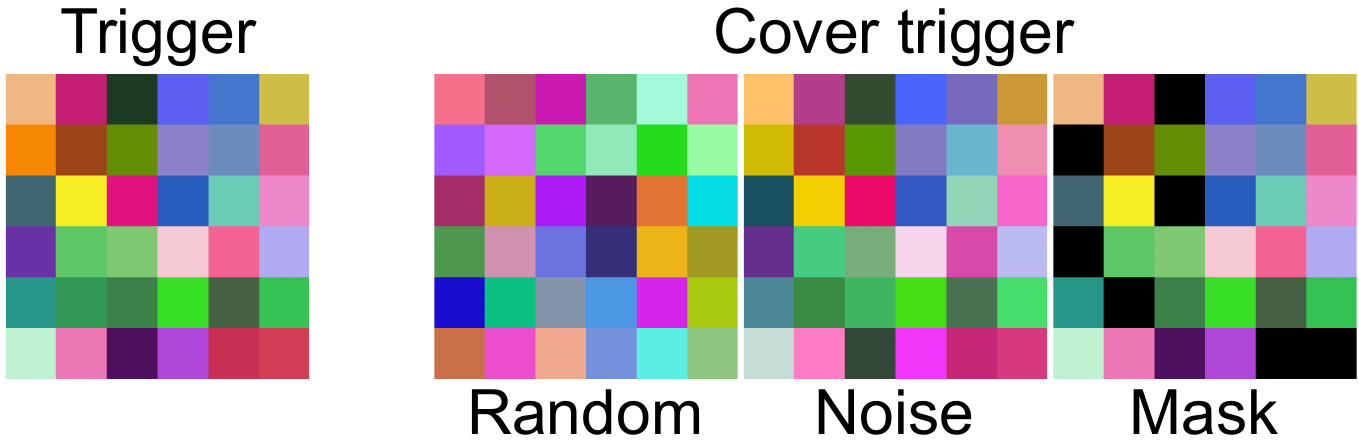}
        \caption{Examples of cover triggers constructed in three different methods.}
    \label{fig:cover_trigger}
\end{figure}

\noindent\textbf{(c) Cover Trigger Construction.} In our previous experiments, we focused on constructing cover triggers (fuzzy triggers carried by cover samples) using a masking-based approach. Here, we explore other possible methods for constructing cover samples. Specifically, we additionally consider two ways to construct cover triggers, namely, random sampling and noise injection.
The random sampling approach constructs a cover trigger by randomly sampling from the distribution of the original trigger. On the other hand, the noise injection approach involves adding noise (Gaussian noise with a mean of 0 and a standard deviation of 0.1) to the original trigger, creating a cover trigger. \autoref{fig:cover_trigger} illustrates examples of cover triggers constructed using these two methods.

We use these two approaches to build two new cover datasets, and each is used to train a Badnet backdoored model. The models use a ResNet18 architecture and the CIFAR-10 dataset. The final backdoor \metricname achieved by these models is 58.43\% and 80.05\%, respectively. In comparison, the cover triggers constructed based on the masking approach show the most significant improvement in \metricname, reaching 84.55\% under the same settings. The improvement in \metricname with cover triggers constructed through random sampling is limited, mainly because it is challenging to construct cover triggers in a controlled manner using this approach. In contrast, cover triggers constructed by adding noise and the masking approach can be controlled by adjusting the noise intensity and masking ratio. However, constructing cover triggers based on adding noise is challenging to determine the appropriate noise intensity. We opt for the masking-based approach, using a default mask rate of 20\%, which achieves good \metricname enhancement in all settings.

\begin{figure}[t]
    \centering
    \includegraphics[trim=0 0 0 0,clip,width=0.45\textwidth]{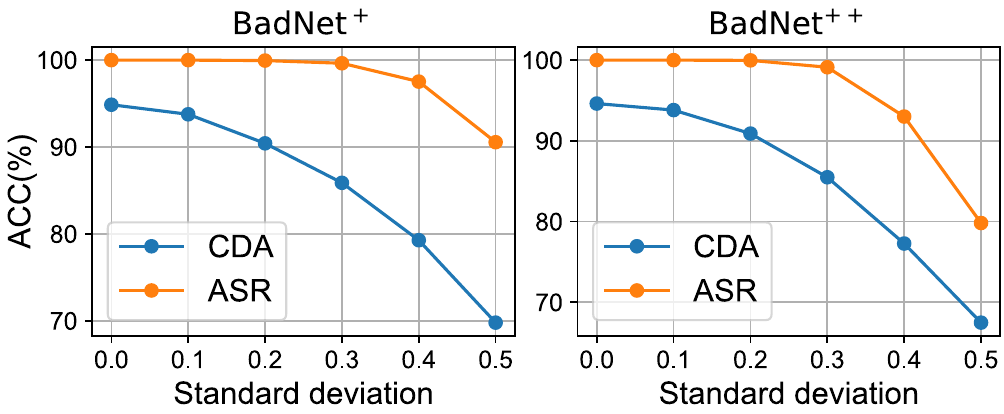}
        \caption{\textcolor{black}{Performance of BELT in Gaussian noise with different standard deviations.}}
    \label{fig:addnoise}
\end{figure}

\textcolor{black}{\noindent\textbf{(d) Adaptive Defense.} In this part, we analyze the potential adaptive defense against BELT. As the effectiveness of BELT relies on the exclusivity of backdoors, only sufficiently precise triggers have the ability to activate the backdoor. In other words, the backdoor exclusivity prevents the backdoor from responding to fuzzy triggers. Therefore, an adaptive defender may attempt to add noises to the input data to make the potential triggers fuzzy, thereby circumventing the backdoor effect. Below, we further verify the resilience of BELT to Gaussian perturbations. As shown in \autoref{fig:addnoise}, for BadNet$^+$ under the CIFAR-10 and ResNet18 settings, CDA/ASR decreases from 94.85\%/99.98\% to 69.79\%/90.54\% as the standard deviation of the noise increases from $0$ to $0.05$. BadNet$^{++}$ exhibits the same trend as BadNet$^+$, with ASR consistently outperforming CDA as the noise intensity increases. This indicates that BELT achieves balance between backdoor exclusivity and perturbation resistance.}

\noindent\textbf{\textcolor{black}{(e)} Limitations and Future Works.}
In this study, we have taken a step towards evaluating and fortifying backdoor \metricname in the context of data outsourcing and model outsourcing scenarios. To precisely quantify backdoor \metricname is computationally challenging. We instead design an optimization-based algorithm to estimate the \metricname score. Our algorithm incorporates optimization directions and dynamic weights to evade the potential local minima during the estimation. The experimental results show the estimated \metricname score does reflect well the stealthiness of different backdoor attacks. Considering the importance of the \metricname score in characterizing the backdoor stealthiness, it would be meaningful for future works to devise more accurate estimation of backdoor \metricname via, .e.g, more advanced numeric techniques.

To evade existing defenses,  \backdoorname suppresses the correlation between the backdoor and the fuzzy triggers, a key trait exploited by the state-of-the-art defenses for backdoor detection. However, in the data outsourcing cases, the activation vectors of the poisoned samples may be outliers to the benign ones for existing backdoor attacks, which may leave an extra dimension for the defender to do, e.g., data cleansing \cite{ma2022beatrix}. Nevertheless, when the attacker has more control over the victim model (e.g., in the model outsourcing case, a more popular scenario due to the rise of third-party model supply chains such as HuggingFace \cite{jiang2022empirical}), \textsc{BELT} makes further efforts in introducing momentum-based center loss to make poisoned sample features to be indistinguishable from benign ones. In our study, we have tried our best to cover the typical backdoor defenses published at top-tier conferences in recent years. Future works may consider further validating our attack technique on more newly proposed backdoor defenses.

\section{Conclusion}\label{sec:conclusion}
In this work, we propose and investigate a novel characteristic of backdoor attacks called \textit{backdoor \metricname}. Based on our proposed approximation algorithm to quantify \metricname, we conduct a comprehensive analysis of the intricate relation between backdoor \metricname and the existence of many fuzzy triggers in the classical backdoor attacks, which are intensively exploited by the state-of-the-art defenses to detect and recover the triggers. To enhance the capability of existing backdoor attacks to evade the state-of-the-art defenses, we introduce \backdoorallname (\backdoorname) which suppresses the association between the backdoor and the potential fuzzy triggers and can be seamless combined with almost any existing backdoor attacks. 
We thoroughly validate the effectiveness of \backdoorname on four existing backdoor attacks which would otherwise be detected by the state-of-the-art defenses.  After applying the BELT technique, the backdoored models exhibit high evasiveness against \textcolor{black}{seven} state-of-the-art backdoor defenses. This is primarily attributed to \backdoorname refining the activation conditions of the backdoor, disrupting the necessary assumptions of these defense mechanisms, and causing them to struggle in obtaining an effective fuzzy trigger in the presence of a strongly exclusive backdoor.


\bibliographystyle{acm}
\bibliography{defs,refs}

\appendices
\section{Trigger Upper Bound Optimization Algorithm}\label{alg}
The detailed optimization process for trigger upper bounds is shown in Algorithm \ref{alg:example}.

\begin{algorithm}
\caption{Optimize Trigger Upper Bound}
\label{alg:example}
\begin{algorithmic}[1]
\Require Backdoored model $F_{\theta'}$, target label $t$, clean image $x$ and mask $m_x$, trigger pattern $p$ and mask $m_p$ , perturbation boundary $\delta_b$
\State Initialize optimizable perturbation $\delta$ and regular term weight $\lambda$
\State Set learning rate $\alpha$ and number of epochs $E$
\For{epoch $\gets 1$ to E}
    \State $x' = \text{clip}(x \cdot m_x + (p + \delta) \cdot m_p, 0, 1)$
    \Comment{Add perturbed trigger for $x$}
    \State $y' = F_{\theta'}(x')$
    
    \If{$epoch > \frac{E}{2}$}
        \State $\widetilde{y'} = \text{normalize}(y')$
        \State $\lambda \leftarrow \lambda + \widetilde{y'_{y_t}} - \underset{i \neq y_t}{\text{max}}(\widetilde{y'_i})$
        \Comment{Dynamic update $\lambda$}
    \EndIf
    
    \State $\mathcal{L} = -y_t \log(y') + \lambda \|\delta_b - \delta\|_2$
    \State $\delta \leftarrow \delta - \alpha \frac{\partial \mathcal{L}}{\partial \delta}$
\EndFor
\State \Return Trigger upper bound $\|\delta\|_2$ for $x$
\end{algorithmic}
\end{algorithm}

\section{Methodology Description}\label{methodology_description}

\subsection{Neural Cleanse}

In addressing the susceptibility of neural networks to backdoor attacks, the Neural Cleanse \cite{wang2019neural} framework focuses on the reverse engineering of triggers, aiming to detect and mitigate the creation of "shortcuts" in the latent space that lead to misclassification. The core concept involves estimating the minimum perturbation necessary to generate the smallest "shortcut" for each targeted category, effectively reverse engineering the trigger. The methodology consists of three key steps: firstly, reverse engineering the smallest trigger for a given label using an optimization algorithm; secondly, iteratively applying this process to generate a one-to-one corresponding reverse-engineered trigger per label; and thirdly, evaluating whether a reverse-engineered trigger for a category is significantly smaller than others, identifying it as a genuine trigger. If no significant outlier is detected, the model is deemed benign. To assess the anomaly index of the model, Neural Cleanse utilizes the Median Absolute Deviation (MAD) for robustness against outliers. The $L_1$ norm of all reverse-engineered triggers is calculated, and MAD is determined by computing the median of absolute deviations. The anomaly index is expressed as the absolute difference between the minimum norm and the median, divided by the product of the median absolute deviation and a constant estimator (1.4826). A model with an anomaly index greater than 2 is considered more than 95\% likely to be backdoored.

\subsection{ABS}

Artificial Brain Stimulation (ABS) \cite{liu2019abs} serves as a backdoor detection methodology, drawing inspiration from Electrical Brain Stimulation (EBS). ABS employs a controlled approach, manipulating the activation of individual artificial neurons while keeping others fixed to assess their impact on classification behavior, thereby scrutinizing neural models for the presence of backdoors. Rooted in the recognition that concealed damaged neurons exert influence over backdoor behavior, ABS leverages Neuron Stimulation Functions (NSF) to gauge the repercussions of neuron activation on label output. ABS initiates the identification of damaged neurons by computing the NSF of the selected layer, designating neurons with significantly heightened output activation as candidate damaged neurons. In an iterative process, ABS scans neurons, marks those with notable output activation increases as damaged candidates, and utilizes an optimized method for reverse engineering to generate triggers for the ten most probable damaged neurons. The determination of a model's susceptibility to backdoors is contingent on a threshold derived from the Reverse-Engineered Trojan Trigger Attack Success Rate (REASR); if the REASR of the reverse trigger surpasses the 88\% threshold, the model is classified as harboring a backdoor, otherwise, it is deemed benign.

\subsection{MNTD}

Meta Neural Trojan Detection (MNTD) \cite{xu2021detecting} focuses on training a meta-classifier capable of distinguishing between backdoored and clean models within a specific dataset. The process involves three main steps: shadow model generation, meta-classifier training, and target model detection. Initially, a multitude of shadow models, comprising both clean and backdoored models, are trained on a small dataset accessible to the defender. These backdoored models undergo various poison attacks, including modification and blending attacks. The resulting shadow models, intentionally trained with limited convergence to minimize training overhead, form the basis for subsequent steps. In the meta-classifier training phase, a modest number of queries are fed into the shadow models, generating concatenated outputs (logits) for each query. These logits, alongside ground-truth labels (benign or backdoored), serve as inputs for training the meta-classifier. Importantly, the queries are iteratively updated to optimize efficiency during meta-classifier training. The final step involves utilizing optimized queries as input for the target model, obtaining logits that are then inputted into the meta-classifier, producing a score. This score serves as an indicator of the target model's malicious degree, with a higher score indicating potential backdoor injection.
MNTD uses the area under the curve (AUC) as the primary metric to evaluate meta-classifier performance. In this case, the meta-classifier first calculates the scores of all models in the test model set and then calculates the AUC based on the model's label (0 for benign models and 1 for backdoored models).

\subsection{MOTH}

Model Orthogonalization (MOTH) \cite{tao2022model} represents an innovative technique for fortifying models, particularly those with potential backdoors, by introducing additional training steps to eliminate any latent vulnerabilities in the model.
To elaborate, MOTH employs a unique approach to compel the model to eradicate low-level backdoor features. This is achieved through an iterative process that generates minimal backdoors between classes $a$ and $b$, as well as between $b$ and $a$. The proposed framework leverages adversarial training to enhance the reinforcement process. MOTH identifies minimal backdoors originating from the victim class towards the target class, utilizing these samples with unaltered labels as training data. This strategic utilization encourages the model to disregard potential target shortcuts and instead prioritize the inherent characteristics of the training samples.
In practical applications, particularly when dealing with models susceptible to hidden backdoors, an initial step involves symmetric hardening to create symmetric backdoors. Subsequently, promising scheduling pairs are carefully chosen based on the maximal increase in the distance between classes. Finally, the model undergoes training in model orthogonalization using these selected pairs.

\subsection{STRIP}

STRIP \cite{gao2019strip} is an online detection method functioning at the data level, is crafted to discern contaminated data harboring triggers within the input. The procedure strategically capitalizes on the input-agnostic strength of triggers, allowing STRIP to identify contaminated inputs through an evaluation of the consistency of perturbed replicas. This method involves the superimposition of the input with clean samples, leading to the creation of multiple copies. These copies, when fed into the model, generate predicted labels, and their consistency is measured through an assessment of entropy. The presence of low entropy in a perturbation signifies a triggering input, whereas high entropy suggests a clean input. STRIP adeptly exploits the dominant hijacking effect of triggers, effectively transforming it into a vulnerability for online detection. In practical terms, as an input is introduced into a backdoored model, the generation of multiple copies is facilitated by overlaying it with clean samples, and the ensuing perturbations undergo scrutiny for their consistency. Predictions characterized by low entropy serve as an indicative marker for a trigger-carrying sample, allowing for a clear differentiation from clean inputs.

\subsection{SentiNet}

SentiNet \cite{chou2020sentinet} is a data-level detection technology encompasses adversarial object localization, the proposal of classes through segmentation, and the generation of masks using GradCAM. Its primary objective of identifying trigger inputs capable of compromising neural network predictions, all without relying on a priori knowledge of the attack vector. The methodology strategically leverages model interpretability and object detection techniques to unveil contiguous regions within input images that wield significant influence over classification outcomes. These identified regions are postulated to possess a heightened probability of harboring triggers that exert an impact on the classification process. Once pinpointed, these influential regions are systematically excised and transposed onto alternative images endowed with ground-truth labels. The efficacy of this process is contingent upon the attainment of sufficiently elevated misclassification rates and confidence levels for the transposed images, thereby designating the excised patch as an adversarial entity housing a concealed backdoor trigger. 
Specifically, SentiNet overlays the continuously influential regions of the test image onto a set of clean test images. These covered images are then fed into the network, and the success rate of deceiving the model, denoted as Foolde, is calculated. Simultaneously, low saliency inert patterns replace the content of the overlapped regions (e.g., Gaussian noise) to compute the average confidence value of inert patterns, denoted as Average confidence. The combination of these two metrics is used to determine whether a sample carries a trigger.

\end{document}